\documentclass[11pt]{scrartcl}

\usepackage[utf8]{inputenc} 
\usepackage{mathtools, amsfonts, amssymb, breqn}
\usepackage{amsthm}

\usepackage[british]{babel}
\usepackage{graphicx, hyperref, tabularx, abstract, makecell,ragged2e}

\usepackage[format=plain]{caption}
\DeclareCaptionFormat{cont}{#1 (cont.)#2#3\par}

\usepackage[outdir=./]{epstopdf}   

\usepackage{cellspace, hhline} 

\usepackage{subfig}
\usepackage{xcolor-solarized}
\usepackage{todonotes}
\usepackage{overpic}

\usepackage{enumitem}
\setdescription{itemsep=5pt,parsep=0pt,leftmargin=1.55cm,font=\normalfont}

\usepackage[]{geometry}
\newcommand{\dif}{{\operatorname{d}}}
\newcommand{\R}{\ensuremath{\mathbb{R}}}
\newcommand{\C}{\ensuremath{\mathbb{C}}}
\newcommand{\matthree}[9]{\left(
	\begin{array}{ccc}
		#1 & #2 & #3 \\
		#4 & #5 & #6 \\  
		#7 & #8 & #9 
	\end{array}
	\right)
}

\numberwithin{equation}{section}
\numberwithin{figure}{section}
\numberwithin{table}{section}
\definecolor{changes}{RGB}{0, 0, 0}
\begin{document}
\author{L. Eigentler and J. A. Sherratt}

\newcommand{\Addresses}{{
		\bigskip
		\footnotesize
		
		L. Eigentler (corresponding author),  Maxwell Institute for Mathematical Sciences, Department of Mathematics, Heriot-Watt University, Edinburgh EH14 4AS, United Kingdom\par\nopagebreak
		\textit{E-mail address}: \texttt{le8@hw.ac.uk}
		
		\medskip
		
		J. A. Sherratt,  Maxwell Institute for Mathematical Sciences, Department of Mathematics, Heriot-Watt University, Edinburgh EH14 4AS, United Kingdom\par\nopagebreak
		\textit{E-mail address}: \texttt{J.A.Sherratt@hw.ac.uk}

}}

\title{Spatial self-organisation enables species coexistence in a model for savanna ecosystems}
\date{}
\maketitle
\Addresses

\begin{abstract}
	The savanna biome is characterised by a continuous vegetation cover, comprised of herbaceous and woody plants. The coexistence of species in arid savannas, where water availability is the main limiting resource for plant growth, provides an apparent contradiction to the classical principle of competitive exclusion. Previous theoretical work using nonspatial models has focussed on the development of an understanding of coexistence mechanisms through the consideration of resource niche separation and ecosystem disturbances. In this paper, we propose that a spatial self-organisation principle, caused by a positive feedback between local vegetation growth and water redistribution, is sufficient for species coexistence in savanna ecosystems. We propose a spatiotemporal ecohydrological model of partial differential equations, based on the Klausmeier reaction-advection-diffusion model for vegetation patterns, to investigate the effects of spatial interactions on species coexistence on sloped terrain. Our results suggest that species coexistence is a possible model outcome, if a balance is kept between the species' average fitness (a measure of a species' competitive abilities in a spatially uniform setting) and their colonisation abilities. Spatial heterogeneities in resource availability are utilised by the superior coloniser (grasses), before it is outcompeted by the species of higher average fitness (trees). A stability analysis of the spatially nonuniform coexistence solutions further suggests that grasses act as ecosystem engineers and facilitate the formation of a continuous tree cover for precipitation levels unable to support a uniform tree density in the absence of a grass species.
\end{abstract}

	Keywords: vegetation patterns; periodic travelling waves; wavetrains; pattern formation; ecosystem engineering; banded vegetation; tiger bush; reaction-advection-diffusion; spectral stability;

\section{Introduction}\label{sec: Multispecies Slope: Intro}

Savannas are characterised by the coexistence of herbaceous vegetation (\textit{grasses}) and woody plant types (\textit{shrubs} and \textit{trees}) \cite{Scholes1993}. They are a dominating feature of many geographical regions worldwide, occupying over one eighth of the global land surface \cite{Scholes1993, Scholes2003}. Savannas stretch across a wide range of different climate zones, and in particular different aridity zones. If the total precipitation volume in savannas is low, they are referred to as water-limited or (semi-)arid savannas \cite{Sankaran2005}.

The coexistence of grass and trees in arid savannas, in which water is the main limiting resource for vegetation growth, has been of particular interest for many decades (see \cite{Yatat2018} for a review of mathematical models on the subject), as it provides an apparent contradiction to the classical competitive exclusion principle, which states that under competition for the same limiting resource only one species can survive (e.g. \cite{Hutchinson1961}). In broad terms, two different mechanisms that facilitate the coexistence of species in savannas have been established using mathematical modelling: resource niche differentiation and environmental disturbances. The former is based on Walter's hypothesis \cite{Walter1971}, which assumes niche differentiation into different root zones. According to this hypothesis, trees have exclusive access to water in deeper soil layers, while grasses are more efficient in their water uptake in the topsoil layer. Early modelling approaches used Walter's hypothesis to provide an explanation for the coexistence of grasses and trees in savannas \cite{Walker1981, Langevelde2003, Walker1982}.

However, empirical studies later suggested that Walter's hypothesis does not always hold in savannas so that it cannot be regarded as a universal mechanism responsible for species coexistence in water-limited ecosystems \cite{Seghieri1995, Mordelet1997,Belsky1994}. Modelling efforts consequently shifted towards other mechanisms, such as disturbances due to fires (e.g. \cite{Staver2011, Beckage2009}), disturbances due to grazing and browsing (e.g. \cite{Scheiter2007, Synodinos2015}), asymmetric competitive effects that trees impose on grass (e.g. \cite{Tilman1994}), different competitive abilities of trees in different life stages \cite{DOnofrio2015, Baudena2010}, or a combination thereof. The main characteristics in which existing models of the savanna biome differ are their representation of the state variables, water dynamics and disturbance occurrences. Many models (e.g. \cite{Touboul2018}) represent the plant state variables as area fraction covers, following the early model by Tilman \cite{Tilman1994}. However, to account for the fact that plant types are typically not mutually exclusive, other modelling frameworks (e.g. \cite{Beckage2011}) characterise plant variables by the plants' biomass per unit area. The model by Tilman and many of its extensions incorporate the plant species' competition for water implicitly, but extensions (e.g. \cite{Accatino2010}) consider water dynamics explicitly in an ecohydrological framework. The occurrence of fire or grazing/browsing disturbances is described either in a probabilistic (e.g. \cite{DOdorico2006}) or a deterministic sense. Models assuming the latter either provide a time-continuous (e.g. \cite{Yu2014}), a time-discrete (\cite{Higgins2010}) or a time-impulsive (\cite{Tchuinte2016, Tchuinte2017, Yatat2017, Yatat2018a}) description of the ecosystem dynamics.

Existing models describing savannas mostly use systems of ordinary differential equations or impulsive differential equations, with the spatiotemporal model for tree cover in mesic savannas by Martinez-Garcia et al. \cite{Martinez-Garcia2013} being a notable exception. Such models are nonspatial and do not take into account any spatial effects that affect the plant populations. However, spatial self-organisation of plants into patterns of alternating patches of high biomass and bare soil are known to be an essential element in the survival of plants in drylands \cite{Deblauwe2008, Valentin1999}. The formation of patterns is usually induced by a positive feedback between local vegetation growth and water redistribution, caused, for example, by the formation of infiltration-inhibiting soil crusts that induce overland water flow towards existing biomass patches \cite{Meron2016, Rietkerk2008}. A very common type of patterned vegetation is stripes that occur on sloped ground (up to 2\% gradient) parallel to the terrain contours \cite{Valentin1999}. Similar to savanna ecosystems, coexistence of trees and grasses (on the level of single vegetation patches) also occurs in patterned vegetation \cite{Herbes2001, Seghieri1997}. In striped vegetation, grass species are usually observed to dominate the uphill region of a stripe, while woody vegetation is more dominant towards the centre and downslope end of a stripe \cite{Seghieri1997, Herbes2001}. 

Spatially explicit mathematical modelling using partial differential equations (PDEs) has explored different mechanisms that enable species coexistence in patterned ecosystems of dryland vegetation. For example, if a pattern-forming species and a non-pattern forming (in the absence of any competitors) species are considered, the pattern-forming species can act as an \textit{ecosystem engineer} by altering the environmental conditions (in particular the availability of water) and thus facilitate coexistence with a non-pattern-forming species superior in its water uptake and dispersal capabilities \cite{Baudena2013, Nathan2013}. A different mechanism that provides a possible explanation for the stability of coexistence patterns is the plant species' adaptation to different soil moisture levels, using the stabilising effect of resource niche differentiation, similar to the early savanna models based on Walter's hypothesis \cite{Callegaro2018, Ursino2006}. Coexistence of species in patterned form may not necessarily be observed as a stable solution of the system, but can also as a long transient, often referred to as a metastable state \cite{Gilad2007a, Eigentler2019Multispecies}. Such metastable patterns occur if the facilitative effects that cause the formation of patterns occur on a much shorter timescale than the competitive effects that yield the eventual extinction of the inferior species. In-phase spatial patterns are not the only context in which coexistence of plant species in patterned form is studied in mathematical models of dryland ecosystems. Alternatively, coexistence of species can occur through the existence of a multitude of localised patterns of one species in an otherwise uniform solution of a competitor (homoclinic snaking) \cite{Kyriazopoulos2014} in a model that assumes a trade-off between root and shoot growth and the associated competition for water and light. 

Most models describing species coexistence in dryland ecosystems are extensions of either the Gilad et al. model \cite{Gilad2004,Gilad2007} or the Klausmeier model \cite{Klausmeier1999}, which are both phenomenological single-species models that capture the formation of vegetation patterns in water-limited ecosystems. The latter in particular stands out due to its deliberately basic description of the plant-water dynamics and thus provides an excellent framework for mathematical analysis and model extensions (e.g. \cite{Bastiaansen2018, Bennett2019, Consolo2019, Consolo2019a, Eigentler2019Multispecies, Eigentler2018nonlocalKlausmeier, Marasco2014, Sherratt2010, Sherratt2005, Sherratt2011, Sherratt2013, Sherratt2013III, Sherratt2013IV, Sherratt2013V, Sherratt2007, Siero2018, Siero2019, Siteur2014,Wang2019, Wang2018a, Ursino2006}). Other modelling frameworks that address the dynamics of vegetation patterns exist (see \cite{Borgogno2009, Martinez-Garcia2018} for reviews), but, to the best of our knowledge, have not been utilised to address species coexistence.

In this paper, we introduce a spatially explicit ecohydrological PDE model to investigate the role of spatial self-organisation principles in the stable coexistence of trees and grasses on sloped terrain in savannas (Sec. \ref{sec: Multispecies Slope: model}). {\color{changes} To solely focus on the effects of spatial heterogeneities caused by a pattern formation feedback, we deliberately assume that both species only differ in their basic parameters, but not in any of their functional responses.} We base our model on the Klausmeier model for vegetation patterns and find stable solutions of the multispecies model in which both species coexist, representing a savanna biome. More precisely, these stable solutions are periodic in space, but, unlike in the single-species Klausmeier model, plant densities in the troughs of the pattern are not close to zero. Instead, both plant densities oscillate between two non-zero values. In Sec. \ref{sec: Multispecies slope: origin of coex pattern} we perform a bifurcation analysis of the model to disentangle the origins of the coexistence state and establish key conditions required for the existence of coexistence patterns. We augment our results on pattern existence by an analysis of their stability in Sec. \ref{sec: Multispecies slope: stability} and address the phase difference between the oscillations in both plant densities in Sec. \ref{sec: Multispecies Slope: phase diff}. Our analysis is restricted to a one-dimensional space domain which is assumed to represent a sloped terrain, as the inclusion of a term describing the flow of water in the downhill direction facilitates the application of a numerical continuation method to study pattern existence and stability. We briefly comment on model solutions on a flat spatial domain in Sec. \ref{sec: Multispecies Slope: Discussion} and discuss the relevance and implications of our results. Sec. \ref{sec: Multispecies Slope: spectrum calc} provides an outline of the numerical continuation methods used in our bifurcation and stability analysis.


\section{The model}\label{sec: Multispecies Slope: model}

In this section, we present the modelling framework used in this paper to study the coexistence of plant species in water-deprived ecosystems. Our model is based on the reaction-advection-diffusion model by Klausmeier \cite{Klausmeier1999}, which in nondimensional form reads
\begin{subequations}\label{eq: Multispecies slope: one species Klausmeier model}
	\begin{align}
	\frac{\partial u }{\partial t} &= \overbrace{u^2w}^{\text{plant growth}} - \overbrace{Bu}^{\text{plant loss}} + \overbrace{\frac{\partial^2 u}{\partial x^2}}^{\text{plant dispersal}}, \\
	\frac{\partial w }{\partial t} &= \underbrace{A}_{\text{rainfall}} - \underbrace{w}_{\substack{\text{evaporation and}\\\text{transpiration}}} - \underbrace{u^2w}_{\substack{\text{water uptake} \\ \text{by plants}}} + \underbrace{\nu\frac{\partial w}{\partial x}}_{\substack{\text{water flow}\\ \text{downhill}}} +  \underbrace{d\frac{\partial^2w}{\partial x^2}}_{\substack{\text{water} \\ \text{diffusion}}}.
	\end{align}
\end{subequations}
The density $u(x,t)$ denotes the dry biomass per unit area, and $w(x,t)$ quantifies the mass of water per unit area at time $t>0$ at a space point $x\in\R$ on a one-dimensional infinite spatial domain, on which $x$ increases in the uphill direction if the terrain is considered to be sloped. It is assumed that rainfall is continuous and that both biomass density and water density decay due to plant mortality and water transpiration and evaporation, respectively, at constant rates. The nonlinearity in the terms describing water consumption by plants and the consequential increase in biomass accounts for part of the positive feedback between local vegetation growth and the redistribution of water. Water uptake is the product of the consumer density ($u$), the resource density ($w$) and a term that accounts for the infiltration of water into soil layers where roots are present ($u$). The latter's dependence on the biomass density stems from the plants' infiltration-enhancing soil modifications and the formation of soil crusts in regions of low biomass. Both densities undergo diffusion and water flow in the downhill direction is modelled by an advection term, if the model is considered on sloped terrain. Diffusion of water was not part of Klausmeier's original model, but is a well-established addition to account for water flow on flat terrain (e.g. \cite{Kealy2012, Siteur2014, Stelt2013, Zelnik2013}). The parameters $A$, $B$, $\nu$ and $d$ are combinations of several dimensional parameters, but represent precipitation, plant mortality rate, the speed of water flow downhill and the ratio of the diffusion coefficients, respectively. 

In a previous paper \cite{Eigentler2019Multispecies}, we have extended the single-species Klausmeier model \eqref{eq: Multispecies slope: one species Klausmeier model} by separating the biomass density $u$ into two species, $u_1$ and $u_2$ with differing growth and mortality rates, diffusion coefficients and water infiltration enhancement strengths. In this paper, we follow a similar approach and analyse the two-species model, which, after a suitable nondimensionalisation (see \cite{Eigentler2019Multispecies}\footnote{The advection parameter $\nu$ is not given in the nondimensionalisation in \cite{Eigentler2019Multispecies}, but $\nu=\tilde{\nu}(k_1k_2)^{-1/2}$, where $\tilde{\nu}$, $k_1$ and $k_2$ are dimensional parameters describing water flow speed, diffusion of species $u_1$ and water evaporation rate, respectively.}), is
\begin{subequations}\label{eq: Multispecies slope: Model: nondimensional model}
	\begin{align}
	\frac{\partial u_1}{\partial t} &= \overbrace{wu_1\left(u_1 + Hu_2\right)}^{\text{plant growth}} - \overbrace{B_1 u_1}^{\substack{\text{plant} \\ \text{mortality}}} + \overbrace{\frac{\partial^2 u_1}{\partial x^2}}^{\text{plant dispersal}}, \\
	\frac{\partial u_2}{\partial t} &=\overbrace{Fwu_2\left(u_1 + Hu_2\right)}^{\text{plant growth}} - \overbrace{B_2 u_2}^{\substack{\text{plant} \\ \text{mortality}}}  +\overbrace{D\frac{\partial^2 u_2}{\partial x^2}}^{\text{plant dispersal}}, \\
	\frac{\partial w}{\partial t} &= \underbrace{A}_{\text{rainfall}}-\underbrace{w}_{\substack{\text{evaporation and}\\\text{transpiration}}} - \underbrace{w\left(u_1+u_2\right)\left(u_1 + Hu_2\right)}_{\text{water uptake by plants}}+ \underbrace{\nu \frac{\partial w}{\partial x}}_{\substack{\text{water flow} \\ \text{ downhill}}} +\underbrace{d \frac{\partial^2 w}{\partial x^2}}_{\substack{\text{water}\\\text{diffusion}}}.
	\end{align}
\end{subequations}
As in \eqref{eq: Multispecies slope: one species Klausmeier model}, $u_i(x,t), i=1,2$ and $w(x,t)$ denote the respective plant densities and the water density at time $t>0$ and point $x\in\R$, where the space coordinate increases in the uphill direction of the sloped terrain. The modelling assumptions are identical to those in the single-species model, i.e. all three densities diffuse, where the nondimensional diffusion coefficients $D$ and $d$ are ratios of the respective dimensional diffusion coefficient and the diffusion coefficient of species $u_1$; water flows downhill; plant loss of both species occurs at constant rates $B_i$; evaporation and transpiration effects reduce the water density at a constant rate; and precipitation continuously supplies the system with water at a constant rate, represented by the nondimensional precipitation parameter $A$. The water uptake term is composed of the total consumer density ($u_1+u_2$), the resource density ($w$), and the enhancement of water infiltration caused by plants ($u_1+Hu_2$). The constant $H$ accounts for the unequally strong effects of different plant species on the soil's permeability. Plant growth of species $u_1$ directly corresponds to the resource consumption by $u_1$ and thus occurs at rate $w(u_1+Hu_2)$. Similarly, the biomass of species $u_2$ increases at rate $Fw(u_1+Hu_2)$, where $F$ is the ratio of the species' water to biomass conversion coefficients. The multispecies model \eqref{eq: Multispecies slope: Model: nondimensional model} is a simple extension of the single-species Klausmeier model \eqref{eq: Multispecies slope: one species Klausmeier model}. The plant species only differ in their parameters, with all functional responses being identical. In particular, each species satisfies the single-species model \eqref{eq: Multispecies slope: one species Klausmeier model} in the absence of its competitor.

{\color{changes} While the multispecies model \eqref{eq: Multispecies slope: Model: nondimensional model} is similar to the model analysed in our previous paper \cite{Eigentler2019Multispecies}, the results presented in this paper address a solution type with applications to a fundamentally different ecosystem.  In \cite{Eigentler2019Multispecies}, we focussed on species coexistence in vegetation patterns, which are characterised by a mosaic of colonised ground and bare soil. In this context, we found that coexistence can occur as a metastable state, that is an inherently unstable state which appears as a long transient in the system. The novelty of the work presented in this paper is twofold. Firstly, we address the effect of spatial interactions on species coexistence in savannas, an ecosystem in which plant cover is continuous, but not necessarily uniform. With the notable exception of \cite{Martinez-Garcia2013}, spatial effects on savanna ecosystems have not been considered in mathematical models before. Secondly, we are able to show that, unlike in the context of patterned vegetation considered in \cite{Eigentler2019Multispecies}, coexistence states of the multispecies model \eqref{eq: Multispecies slope: Model: nondimensional model} that represent a savanna biome are stable solutions.}

The model introduced in \cite{Eigentler2019Multispecies} further includes an asymmetric direct competition term through which one species increases the mortality rate of its competitor (e.g. due to shading). {\color{changes} However, the inclusion of such a direct competition term in either or both of the equations does not yield any qualitative differences in the results on species coexistence presented in this paper (but may, in general, add to the richness of solution types in the system). Quantitative effects of direct interspecific competition include changes to the notion of the \textit{local average fitness} of a species, but in the interest of providing a basic representation of the self-organisation principle as a coexistence mechanism, we do not consider any direct interaction between the plant species in \eqref{eq: Multispecies slope: Model: nondimensional model}. Instead, the two plant species only compete indirectly through the depletion of the limiting resource. }

The main focus of this paper is a description of coexistence of grass and trees or shrubs in water-deprived ecosystems. Thus, we henceforth consider $u_1$ to be a herbaceous species and $u_2$ to be of woody type. {\color{changes}This assumption allows for qualitative statements on the parameters in the system. For example, mortality rates can be inferred from the lifespan of a species. The difference in the typical lifespans of grasses and trees yields that grasses die at a faster rate ($B_1>B_2$) \cite{Accatino2010}. Similarly, plant growth parameters can be deduced from the time necessary for a plant population to reach its equilibrium density. Grasses require significantly shorter periods to attain steady steady state biomass levels than trees, which suggests that grasses are superior in their water-to-biomass conversion ($F<1$) \cite{Accatino2010}. If other system parameters are known, the strength of a plant species' enhancement of water infiltration into the soil can be estimated from its equilibrium density \cite{Klausmeier1999}. As steady state biomass densities for tree species are in general much higher than those of grass species in dryland ecosystems, this yields that grasses cause a larger increase in soil permeability per unit biomass than trees ($H<1$) \cite{Mauchamp1994}. The plant species' diffusion coefficients relate the spatial spread of vegetation with time. The longer generation time of trees suggests slower dispersal of trees ($D<1$).
	
	All our parameter estimates are based on previous modelling studies (e.g. \cite{Klausmeier1999, Siteur2014}), as there is a lack of empirical data that would allow for an accurate parameter estimation. However, all our assumptions on parameter differences between tree and grass species are in agreement with parameter estimates in previous multispecies models (e.g. \cite{Baudena2013, Gilad2007}). Unless otherwise stated, we set $B_1 = 0.45$, $B_2 = 0.0486$, $D=0.109$, $F=0.109$, $H=0.109$, $\nu=50$ and $d=500$.
}

\section{Existence and onset of patterns in which species coexistence occurs} \label{sec: Multispecies slope: origin of coex pattern}

In this section, we discuss the existence of solutions of \eqref{eq: Multispecies slope: Model: nondimensional model} in which both species coexist. Such solutions are periodic travelling waves, i.e. spatially periodic solutions that move in the uphill direction of the domain at a constant speed. Numerical continuation shows that the branches of periodic travelling waves, in which both plant species are strictly positive, terminate at a single-species pattern at either end. The key ingredient in understanding the onset and existence of coexistence states is information on the single-species patterns' stability. An investigation of the essential spectrum of the single-species pattern reveals that bifurcations to coexistence states occur as a single-species pattern loses/gains stability to the introduction of its competitor.

\subsection{Stability of spatially uniform equilibria}
The starting point of our bifurcation analysis is the equilibrium states in a spatially uniform setting. Depending on the level of precipitation, the multispecies model \eqref{eq: Multispecies slope: Model: nondimensional model} has up to five spatially uniform steady states: a trivial desert steady state $(0,0,\overline{w}^D) = (0,0,A)$ which exists and is stable in the whole parameter space; a pair of single-species grass equilibria $(\overline{u}_1^{G,\pm}, 0, \overline{w}^{G,\pm})$ that exist for sufficiently high rainfall volumes $A>A_{\min}^G$; and a pair of single-species tree states $(0,\overline{u}_2^{T,\pm},  \overline{w}^{T,\pm})$ that exist for $A>A_{\min}^T$. In both cases, the pair of single-species equilibria meet in a fold at their respective existence thresholds, and the lower branches, here denoted by a minus sign in the superscripts, are unstable. The remaining single-species grass equilibrium $(\overline{u}_1^{G,+}, 0, \overline{w}^{G,+})$ is linearly stable to spatially uniform perturbations if $B_2-FB_1 >0$ and $B_1<2$, while the tree steady state $(0,\overline{u}_2^{T,+},  \overline{w}^{T,+})$ is linearly stable to spatially homogeneous perturbations if $B_2-FB_1 <0$ and $B_2<2$. \cite{Eigentler2019Multispecies}. Parameter estimates consistently imply that plant mortality is sufficiently low to assume $B_i<2, i=1,2$.

These two stability criteria emphasise the critical role of the quantity $B_2-FB_1$ in the system, as $B_2-FB_1 = 0$ is a separatrix of the stability regions of the single species equilibria in the spatially uniform setting. We thus refer to $B_2-FB_1$ as the \textit{average fitness difference} between the two species, because its sign determines the single-species state to which the system converges in the absence of any spatial interactions (provided the precipitation level $A$ is sufficiently high). In dimensional parameters, the average fitness of a species in the model is the ratio between its water-to-biomass conversion capabilities (growth rate) and its mortality rate \cite{Eigentler2019Multispecies}.

\subsection{Single-species patterns}

If spatial interactions are included, the multispecies model \eqref{eq: Multispecies slope: Model: nondimensional model} admits single-species patterns that move in the uphill direction of the domain at a constant speed. Such regularly patterned solutions moving through the spatial domain are classified as periodic travelling waves, an important solution type for reaction-advection-diffusion equations and other partial differential equations. Periodic travelling waves can be represented by a single variable $z=x-ct$ only, where $c\in\R$ is the migration speed of the periodic solution, and $u_1(x,t) = U_1(z)$, $u_2(x,t) = U_2(z)$ and $w(x,t) = W(z)$. This coordinate transformation reduces the PDE system \eqref{eq: Multispecies slope: Model: nondimensional model} to the corresponding travelling wave ODE system
\begin{subequations}\label{eq: Multispecies slope: Model: tw model}
	\begin{align}
	WU_1\left(U_1 + HU_2\right) - B_1 U_1 + c\frac{\dif U_1}{\dif z} +\frac{\dif^2 U_1}{\dif z^2} &=0,\label{eq: Multispecies: Model: tw model u1} \\
	FWU_2\left(U_1 + HU_2\right) - B_2 U_2 + \frac{\dif U_2}{\dif z} +D\frac{\dif^2 U_2}{\dif z^2}&=0, \label{eq: Multispecies: Model: tw model u2}\\
	A-W - W\left(U_1+U_2\right)\left(U_1 + HU_2\right) + (c+\nu) \frac{\dif W}{\dif z} +d \frac{\dif^2 W}{\dif z^2} &=0.
	\end{align}
\end{subequations}

Patterned solutions of the PDE system \eqref{eq: Multispecies slope: Model: nondimensional model} correspond to limit cycles of \eqref{eq: Multispecies slope: Model: tw model}. In the PDE setting of \eqref{eq: Multispecies slope: Model: nondimensional model}, we would typically investigate the interval of a given control parameter, here the precipitation parameter $A$, in which patterned solution exist. Moreover, the transformation to the comoving frame introduces an additional parameter: the migration speed $c$. If a patterned solution of \eqref{eq: Multispecies slope: Model: nondimensional model} exist for a given set of the PDE parameters, limit cycles of \eqref{eq: Multispecies slope: Model: tw model} exist for a range of values of the migration speed $c$. We thus need to consider a pattern forming region in the $(A,c)$ parameter plane, instead of an interval of $A$ only.

The existence of single-species patterns is examined using the numerical continuation software AUTO-07p \cite{AUTO} and form part of the bifurcation diagrams visualised in Fig. \ref{fig: Multispecies slope: bifurcation diag}. In particular, since the multispecies model \eqref{eq: Multispecies slope: Model: nondimensional model} reduces to the single-species Klausmeier model \eqref{eq: Multispecies slope: one species Klausmeier model} in the absence of one of the species, the bifurcation structure of the system's single-species states is identical to that of the single-species Klausmeier model.  More precisely, the pair of spatially uniform single-species grass equilibria $(\overline{u_1}^{G,\pm},0,\overline{w}^{G,\pm})$ meet in a fold. In the spatial model, the branch stable to spatially uniform perturbations loses its stability at a Turing-Hopf bifurcation. This is the onset locus of the single-species pattern. A multitude of stable and unstable patterned states at different wavelengths and migration speeds exist (only one solution branch is shown in the bifurcation diagrams \ref{fig: Multispecies slope: bifurcation diag}), which all originate at a Hopf-bifurcation and terminate in a homoclinic orbit as the control parameter $A$ is decreased \cite{Sherratt2007}. Due to the symmetry in the model, identical considerations hold true for the single-species tree states.

\subsection{Multispecies patterns}

\begin{figure}
	\centering
	\includegraphics[width=\textwidth]{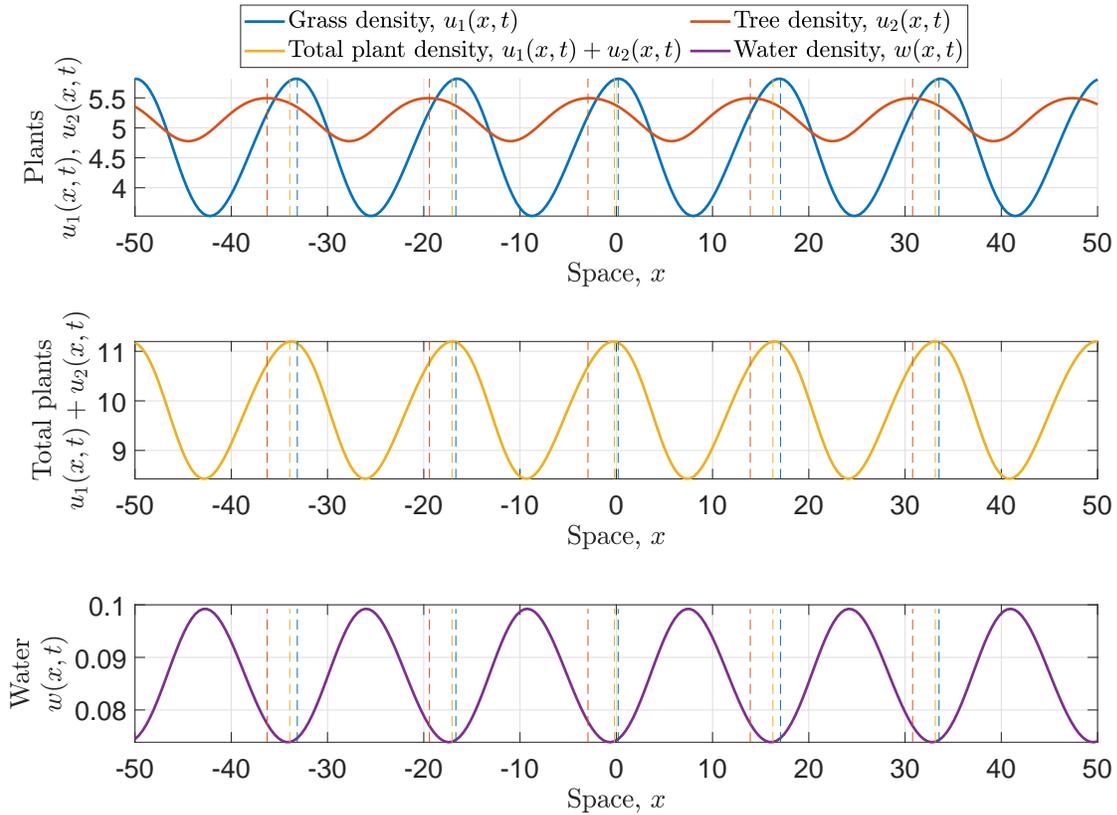}
	\caption{Numerical simulation of the multispecies model. This figure shows a typical patterned solution of \eqref{eq: Multispecies slope: Model: nondimensional model} in which both species coexist. The red, blue and yellow vertical lines indicate the location of local minima of the grass density $u_1$, the tree density $u_2$ and the total plant density $u_1+u_2$ respectively, and highlight {\color{changes}that the total plant density and the water density are antiphase, as well as the existence of a phase difference between the plant patterns}. The solution is obtained through a numerical simulations with precipitation parameter $A=4.5$.}\label{fig: Multispecies slope: numerical simulation}
\end{figure}

Even though there is no spatially uniform equilibrium in which both plant species coexist, numerical simulations of the full system (Fig. \ref{fig: Multispecies slope: numerical simulation}) suggest the existence of stable patterned solutions of \eqref{eq: Multispecies slope: Model: nondimensional model} in which species coexistence occurs. Such solutions also move in the uphill direction, but are distinctly different from the single-species patterns that occur in both the single-species Klausmeier model \eqref{eq: Multispecies slope: one species Klausmeier model} and the multispecies model \eqref{eq: Multispecies slope: Model: nondimensional model}. In single-species patterned solutions, the plant density oscillates between a high level of biomass and a biomass level close to zero (Fig. \ref{fig: Multispecies Slope: wavelength compare sol plots} (a) and (b)). Ecologically, such solutions represent a transect of a striped vegetation pattern in which patches of high biomass alternate with regions of bare soil. By contrast, in the multispecies patterns, both plant densities oscillate between two nonzero levels (Fig. \ref{fig: Multispecies slope: numerical simulation} and Fig. \ref{fig: Multispecies Slope: wavelength compare sol plots} (c) and (d)). In this solution type, there are no patches devoid of biomass, as occurs in a savanna ecosystem. 

\subsubsection{Onset of multispecies patterns}
Branches of single-species periodic travelling waves originate from bifurcations of the spatially uniform equilibria. Further bifurcations may occur along those solution branches, and these are the origin of other solution branches in which both plant species coexist (with non-negative densities) in a patterned state. An insight into the onset of these coexistence patterns is gained through a stability analysis of the single-species patterns in both the single-species Klausmeier model \eqref{eq: Multispecies slope: one species Klausmeier model} and the multispecies model \eqref{eq: Multispecies slope: Model: nondimensional model}. The stability of a periodic travelling wave can be determined through a calculation of its essential spectrum. 

The essential spectrum $\mathcal{S} \subset \C$ of a periodic travelling wave solution determines the leading order behaviour of small perturbations to the periodic travelling wave. Since periodic travelling waves are translation invariant, the origin is always part of the essential spectrum. Hence, the origin is excluded from the following definition of stability. If the essential spectrum lies entirely in the $\Re(\lambda)<0, \lambda \in \C$ half-plane, then the periodic travelling wave is spectrally stable, otherwise it is spectrally unstable. The essential spectrum can be calculated using the numerical continuation method by Rademacher et al. \cite{Rademacher2007} and we provide a brief outline of how the method is applied to \eqref{eq: Multispecies slope: Model: tw model} in Sec. \ref{sec: Multispecies Slope: spectrum calc}. 

\begin{figure}
	\centering
	\subfloat[single-species model \label{fig: Multispecies slope: one species pattern spectra comparison one}]{\includegraphics[width=0.48\textwidth]{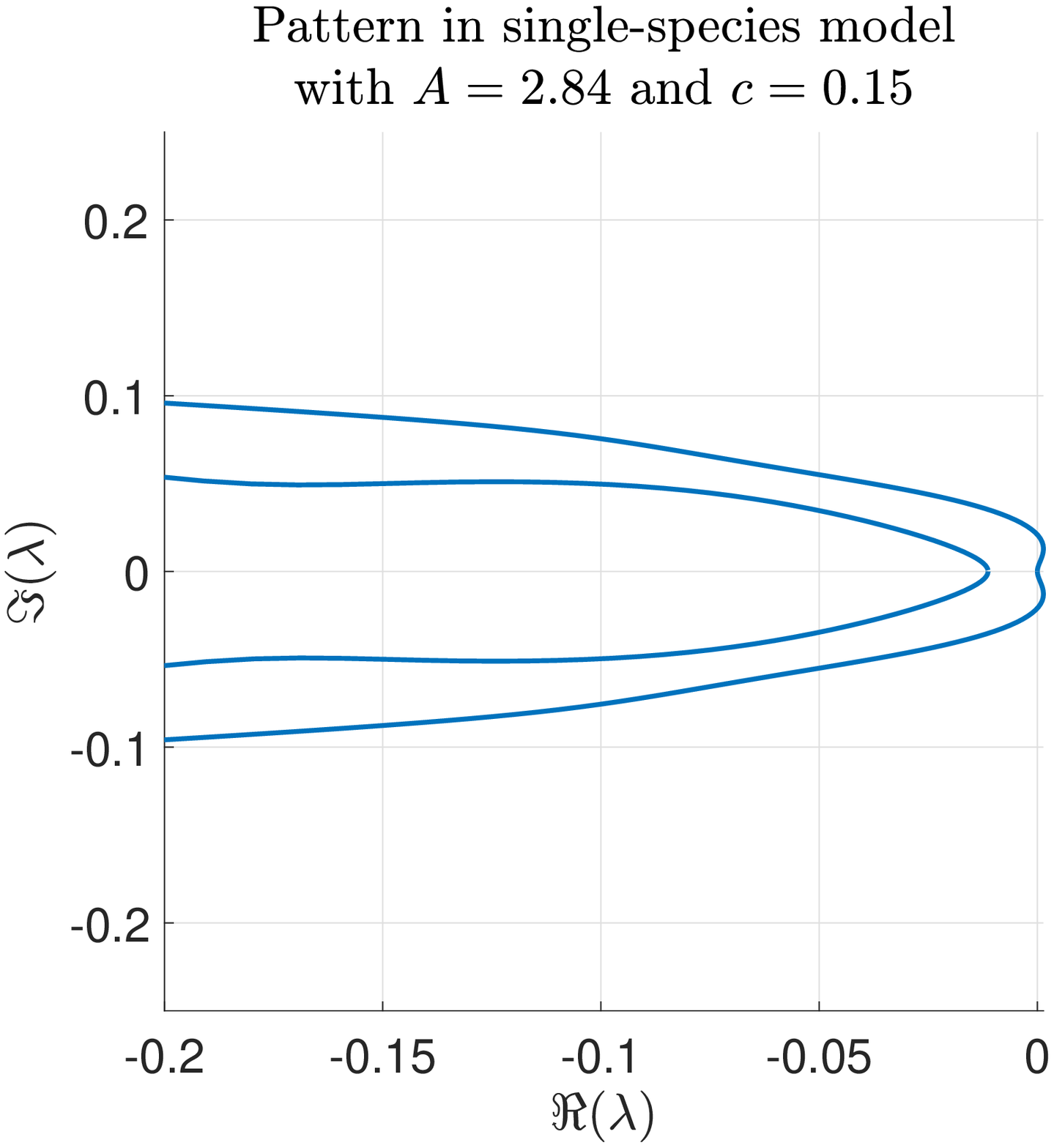}}
	\subfloat[Multispecies model \label{fig: Multispecies slope: one species pattern spectra comparison multi}]{\includegraphics[width=0.48\textwidth]{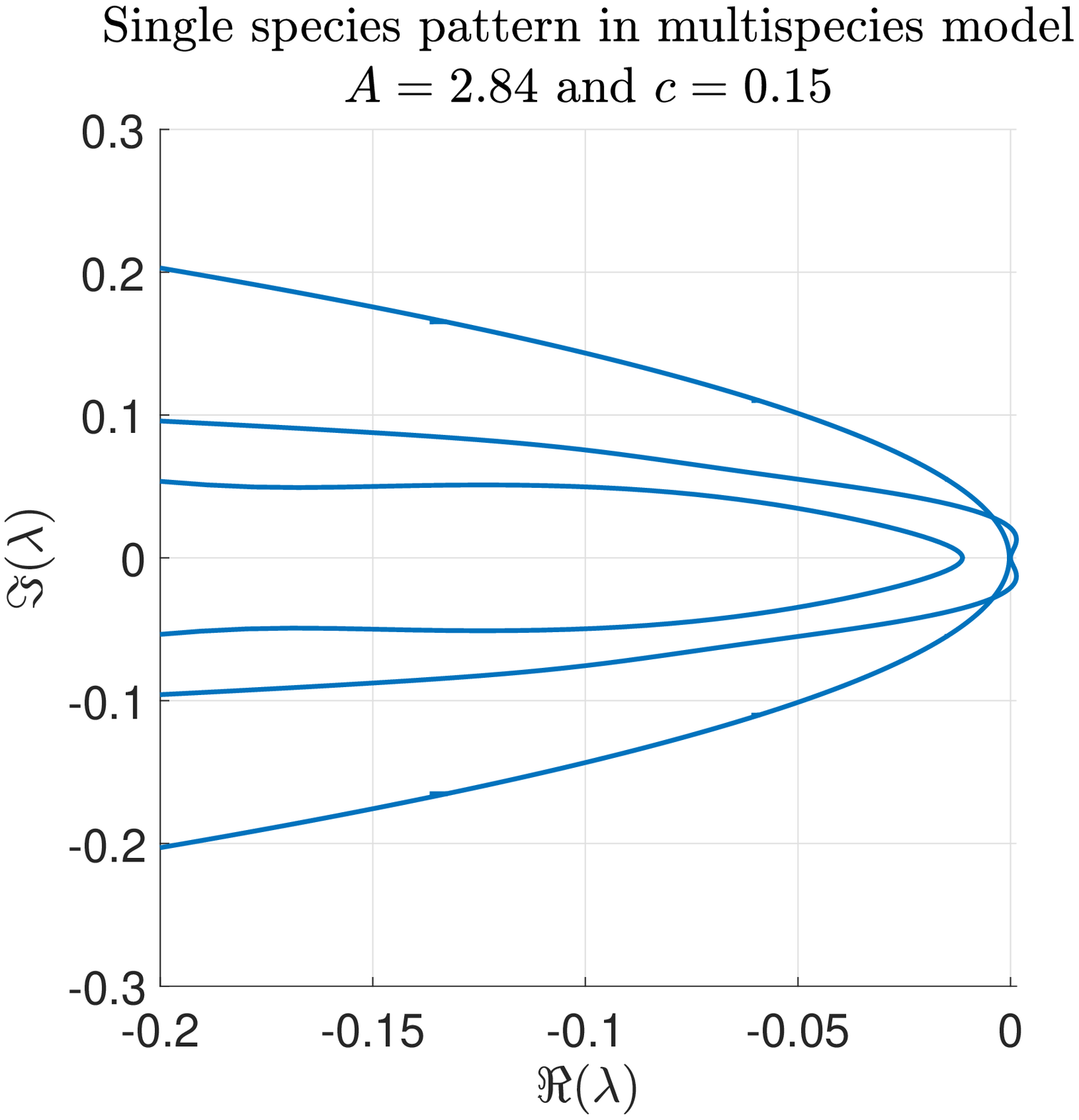}}\\
	\subfloat[Multispecies model \label{fig: Multispecies slope: one species pattern spectra comparison coex pattern bif}]{\includegraphics[width=0.48\textwidth]{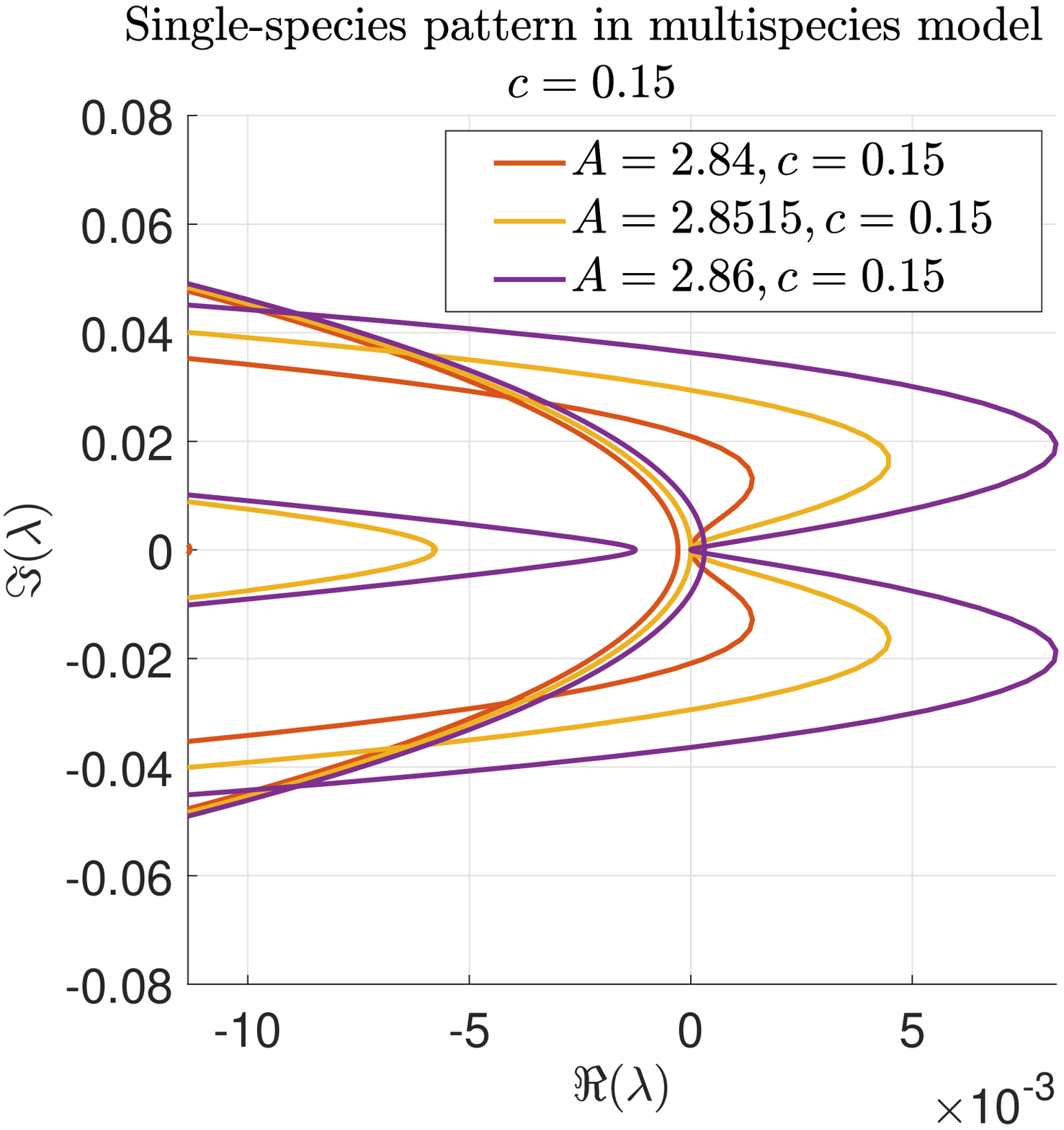}}
	\caption{Spectra of single-species patterns. The visualisations in (a) and (b) compare the spectrum of a patterned solution in the single-species Klausmeier model to that of the identical periodic travelling wave in the multispecies model. The pattern's spectrum in the single-species model is a subset of its the pattern's spectrum in the multispecies model, as the latter contains additional components corresponding to perturbations in the plant density absent in the single species pattern. In (c), the spectra of a single-species pattern in the multispecies model is shown around the origin for different values of the precipitation parameter $A$ (either side of and at the bifurcation to the multispecies pattern) to visualise that the bifurcation to coexistence patterns occurs as the single-species loses/gains stability to the introduction of a second species.}\label{fig: Multispecies slope: one species pattern spectra comparison}
\end{figure}

To understand the onset of coexistence patterns, the essential spectrum of a given pattern in the single-species Klausmeier model \eqref{eq: Multispecies slope: one species Klausmeier model} is compared with that of the same single-species solution of the multispecies model \eqref{eq: Multispecies slope: Model: nondimensional model} (Fig. \ref{fig: Multispecies slope: one species pattern spectra comparison}). The spectrum of the pattern in the multispecies model includes additional components that describe the behaviour of perturbations in the plant type absent in the single species pattern. The bifurcation to the coexistence patterns occurs where the single species pattern loses stability to the introduction of the competitor species. This does not necessarily correspond to a stability change of the single species pattern, since it may be unstable in the single-species model either side of the bifurcation. In more formal words, if $\mathcal{S}_1$ denotes the spectrum of a single-species pattern in the single-species model \eqref{eq: Multispecies slope: one species Klausmeier model} (Fig. \ref{fig: Multispecies slope: one species pattern spectra comparison one}) and $\mathcal{S}_2$ denotes the spectrum of the same solution in the multispecies model \eqref{eq: Multispecies slope: Model: nondimensional model} (Fig. \ref{fig: Multispecies slope: one species pattern spectra comparison multi}), then $\mathcal{S}_1 \subset \mathcal{S}_2$ and the bifurcation to the coexistence pattern occurs as $\max\{\Re(\lambda): \lambda \in \mathcal{S}_2 \setminus \mathcal{S}_1\} = 0$, i.e. as $\mathcal{S}_2 \setminus \mathcal{S}_1$ crosses the imaginary axis $\Re(\lambda)=0$ (Fig. \ref{fig: Multispecies slope: one species pattern spectra comparison coex pattern bif}). Due to the symmetry in the model, these considerations hold for both species in the model.

The coexistence solution branches either connect both single-species solution branches or connect two bifurcations along the same single-species pattern branch. However, coexistence patterns do not originate or terminate at these bifurcations. Instead, the plant density which is zero at the bifurcation changes its sign and the coexistence solution branch continues beyond the bifurcation but is biologically irrelevant (not shown in Fig. \ref{fig: Multispecies slope: bifurcation diag}). We henceforth use \textit{coexistence pattern} to describe those with positive densities in both species only, and with a slight abuse of terminology refer to the branching points along the single species pattern solution branches as their \textit{origins} or \textit{termini}. The exception to the considerations detailed above is large migration speeds $c$, for which only one of the single-species pattern exists. In this case, the branch of patterned coexistence solutions terminates in a homoclinic orbit.

\subsubsection{Existence of multispecies patterns}
A critical requirement for the existence of coexistence patterns is a sufficiently slow (compared to its competitor) growth rate of the species with superior average fitness. If $B_2 - FB_1<0$ ($u_2$ has higher average fitness) then coexistence patterns only occur if $F$ is below a critical threshold $F_{\operatorname{exist}}$. A second significant change of the bifurcation structure occurs at $F = F_{\operatorname{split}}<F_{\operatorname{exist}}$, at which the precipitation interval in which coexistence patterns occur is split into two disjoint intervals. Assuming that the average fitness difference $B_2-FB_1$ and the migration speed $c$ are kept constant, changes to the system's bifurcation structure under increases in $F$ (and associated decreases in $B_2$) can be characterised as follows (Fig. \ref{fig: Multispecies slope: bifurcation diag}): 
\begin{description}
	\item[$F \ll F_{\operatorname{split}}$:] For sufficiently small $F$, there is only one branch of periodic travelling waves in which both species coexist, which connects branching points on either branch of the single species patterns (Fig. \ref{fig: Multispecies slope: bifurcation diag} (a)). 
	\item[$F \approx F_{\operatorname{split}}$ and $F<F_{\operatorname{split}}$:] As the growth rate ratio $F$ is gradually increased, a second pair of branching points moves along each of the single species pattern branches from the homoclinic solution towards the Turing-Hopf bifurcation and a second branch of coexistence patterns connects both branching points (Fig. \ref{fig: Multispecies slope: bifurcation diag} (b)). 
	\item[$F_{\operatorname{split}}<F<F_{\operatorname{exist}}$:] A further increase of $F$ causes a significant change in the bifurcation structure. At the critical threshold $F=F_{\operatorname{split}}$ both coexistence solution branches coincide for some precipitation level. For $F>F_{\operatorname{split}}$ the origins and termini of the solution branches are exchanged and each solution branch connects both branching points on the same single species pattern branch (Fig. \ref{fig: Multispecies slope: bifurcation diag} (c)). This breaks up the existence interval of the coexistence solutions into the union of two disjoint intervals. 
	\item[$F\approx F_{\operatorname{exist}}$ and $F<F_{\operatorname{exist}}$:] Further increases of $F$ increase the gap between the existence intervals and consequently reduce the size of the existence region (Fig. \ref{fig: Multispecies slope: bifurcation diag} (d)). Increases in $F$ also reduce the distance between both branching points along the single species branch, until they meet in a fold at a threshold $F=F_{\operatorname{exist}}^{(i)}$, $i=1,2$, where $F_{\operatorname{exist}}^{(1)}$ and $F_{\operatorname{exist}}^{(2)}$ may differ and depend on other parameters in the model, in particular the diffusion rate ratio $D$.

	\item[$F>F_{\operatorname{exist}}$:] For $F>F_{\operatorname{exist}}^{(i)}$, no branching points along the respective single species pattern branch exist. For the species of inferior average fitness ($u_1$) this is due to the instability of the single-species pattern to the introduction of the second species $u_2$ caused by the combination of the competitor's higher average fitness and sufficiently fast growth rate. In terms of the essential spectrum, this is characterised by the subset $\mathcal{S}_2 \setminus \mathcal{S}_1$ of the essential spectrum of the single-species pattern, which always extends into the $\Re(\lambda)>0$ half-plane, i.e. $\max\{\Re(\lambda): \lambda \in \mathcal{S}_2 \setminus \mathcal{S}_1\} >0$ along the whole solution branch if $F>F_{\operatorname{exist}}^{(1)}$. Vice versa, $\max\{\Re(\lambda): \lambda \in \mathcal{S}_2 \setminus \mathcal{S}_1\} <0$ for the species of higher average fitness ($u_2$) along the branch of single species pattern, if $F>F_{\operatorname{exist}}^{(2)}$, corresponding to the pattern's stability to the introduction of $u_1$. Thus, patterned solutions in which both species coexist cease to occur at $F=F_{\operatorname{exist}}:=\max\{F_{\operatorname{exist}}^{(i)}\}$. The level of $F_{\operatorname{exist}}$ depends on the dispersal behaviour of both plant species and increases monotonically with $|\log(D)|$. In particular, if $D=1$, i.e. the species' diffusion coefficients are equal, $F_{\operatorname{exist}}=F_{\operatorname{exist}}^{(1)}=F_{\operatorname{exist}}^{(2)}=1$ and coexistence patterns cease to occur if both species growth rates are equal. 
\end{description}
\begin{figure}
	\centering
	\begin{overpic}[height=0.94\textheight]{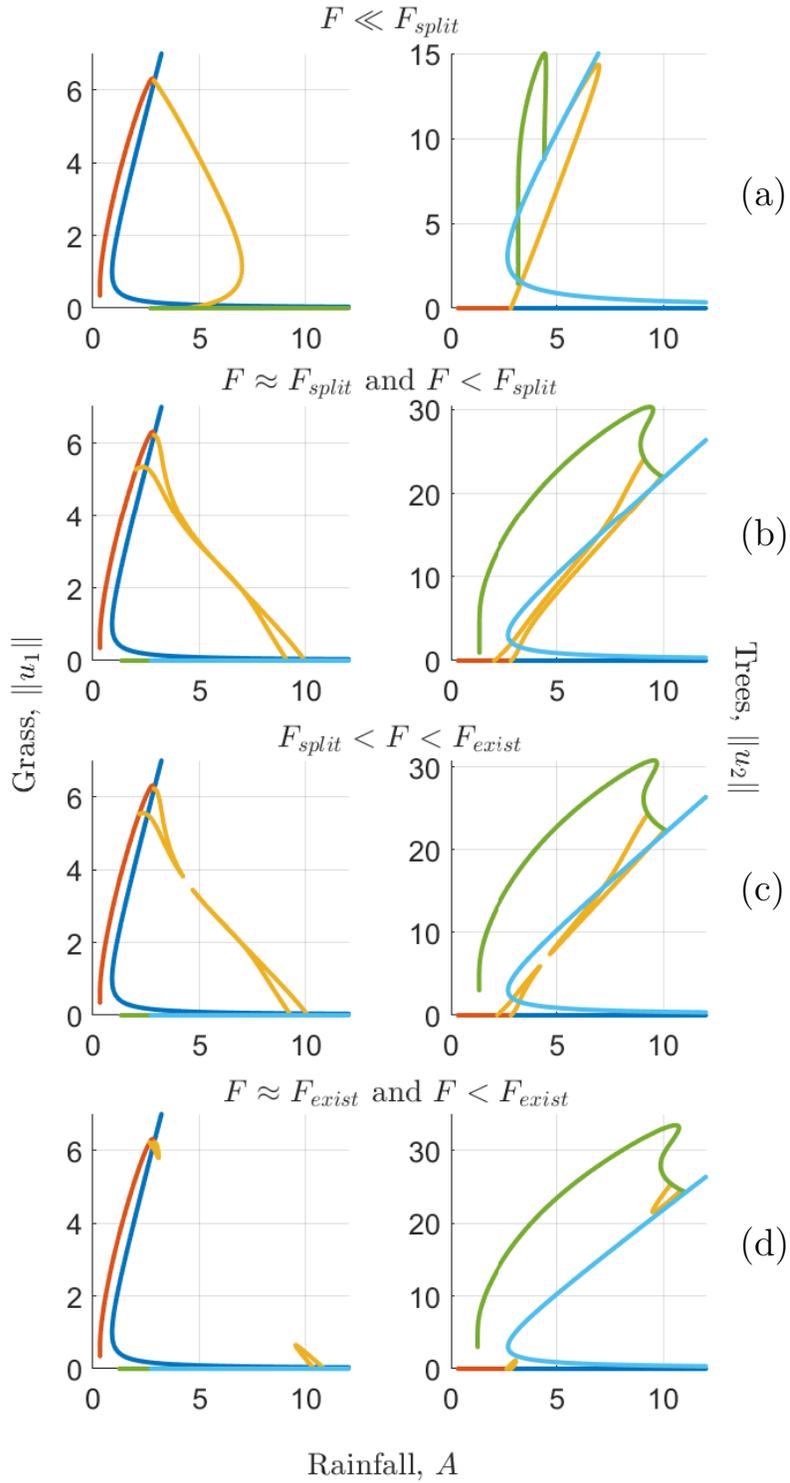}    
		\put(47,83){\Large(a)}   
		\put(47,62){\Large(b)} 
		\put(47,40){\Large(c)} 
		\put(47,18){\Large(d)}                             
	\end{overpic} 
	\caption{Bifurcation diagrams under varying growth rate ratio $F$ and constant average fitness. The full figure caption and legend are displayed overleaf.}\label{fig: Multispecies slope: bifurcation diag}
\end{figure}

\begin{figure}
	\ContinuedFloat
	\captionsetup{list=off,format=cont}
	\begin{minipage}{0.3\textwidth}
		\includegraphics[width=\textwidth]{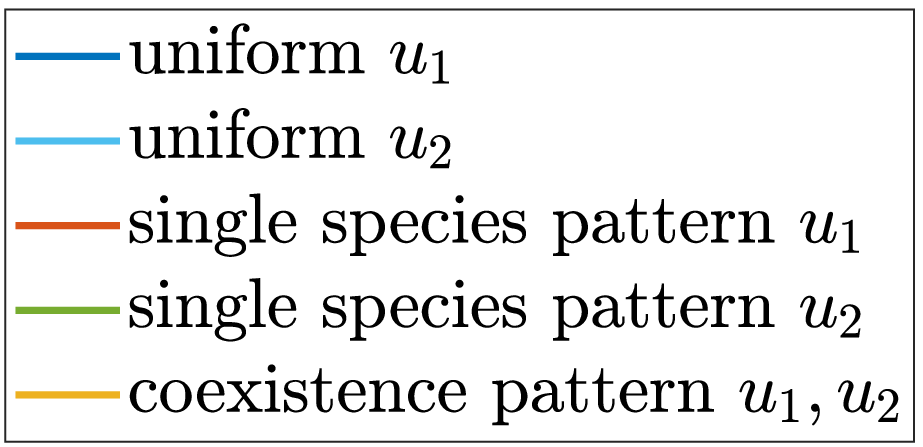}
	\end{minipage}
	\hfill
	\begin{minipage}{0.68\textwidth}
		\caption{Overleaf, bifurcation diagrams for a number of different values of $F$ and $B_2$, keeping the average fitness difference $B_2-FB_1<0$ constant, are shown. For sufficiently small $F$, i.e. a sufficiently slow growth rate of the species of higher average fitness, only one branch of coexistence patterns occurs (a). Increases in $F$ cause the appearance of a second branch (b), before the the precipitation interval in which patterns exist is split into two (c). Further increases of $F$ reduce the size of the parameter region in which coexistence patterns occur (d), before the coexistence state ceases to exist as $F$ passes through a critical threshold (not shown). Solution branches of patterned states are only shown for fixed migration speed $c=0.15$ and no stability information is shown. The chosen values of the growth rate ratio $F$ are $F=0.109$ (in (a)), $F=0.73$ (in (b)), $F=0.7543$ (in (c)) and $F=0.9$ (in (d)). Note the difference to the bifurcation diagrams presented in Fig. \ref{fig: Multispecies Slope: solution under chagnes to avg fitness difference}, in which only $B_2$ is varied and the average fitness difference undergoes changes.}
	\end{minipage}	
\end{figure}

\begin{figure}
	\centering
	\begin{overpic}[height=0.6\textheight]{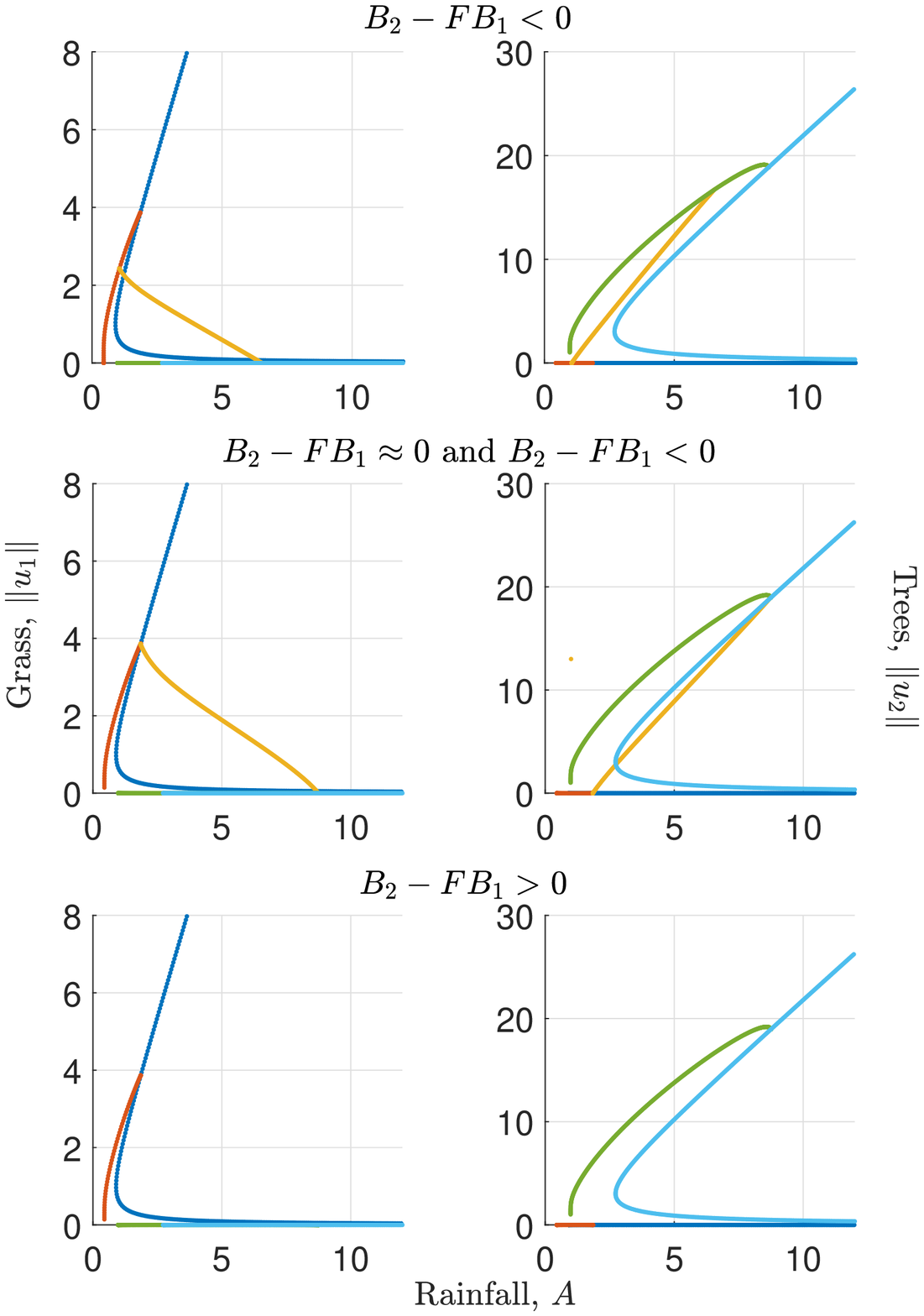}
		\put(70,80){\Large(a)}   
		\put(70,51){\Large(b)} 
		\put(70,20){\Large(c)}	
	\end{overpic}
		\caption{Bifurcation diagrams under changing average fitness difference. Bifurcation diagrams for different values of the average fitness difference $B_2-FB_1$ are shown. As the average fitness difference increases, the origin of coexistence patterns moves along the single species pattern branches towards the Hopf bifurcation at which the single-species pattern originate. No coexistence pattern occur for $B_2-FB_1>0$. The average fitness difference is varied by changes in $B_2$. Plant mortality of the tree species is $B_2 = 0.0486$ (in (a)), $B_2=0.04904$ (in (b)) and $B_2=0.04906$ (in (c)). The legend of Fig. \ref{fig: Multispecies slope: bifurcation diag} applies. Note the difference to the bifurcation diagrams shown in Fig. \ref{fig: Multispecies slope: bifurcation diag}, in which both $F$ and $B_2$ are varied to keep the average fitness difference constant.}\label{fig: Multispecies slope: bifurcation diag avg fitness} 
\end{figure}

The crucial role of the balance between the average fitness difference $B_2-FB_1$ and the growth rate ratio $F$ is further emphasised by an analysis of the bifurcation structure under changes to the average fitness difference if the growth rate ratio $F$ is fixed. If $B_2-FB_1<0$ and $F$ is sufficiently small, i.e. $u_2$ has superior average fitness but a slower growth rate than $u_1$, then coexistence pattern occur, as outlined above (Fig. \ref{fig: Multispecies slope: bifurcation diag avg fitness} (a)). If the average fitness is gradually increased, the branching points, at which the coexistence patterns originate, move along the single species branch towards the Turing-Hopf bifurcation and cease to exist at $B_2-FB_1=0$ (Fig. \ref{fig: Multispecies slope: bifurcation diag avg fitness} (b)). Hence, no coexistence patterns occur if the faster growing species has superior average fitness (Fig. \ref{fig: Multispecies slope: bifurcation diag avg fitness} (c)). In terms of the essential spectrum of the single-species pattern, this is because $\mathcal{S}_2 \setminus \mathcal{S}_1$ does not extend into the $\Re(\lambda)>0$ half-plane for any precipitation levels. This corresponds to the pattern's stability to the introduction of a competitor with slower growth rate and inferior average fitness. 

\begin{figure}
	\centering
	\includegraphics[width=\textwidth]{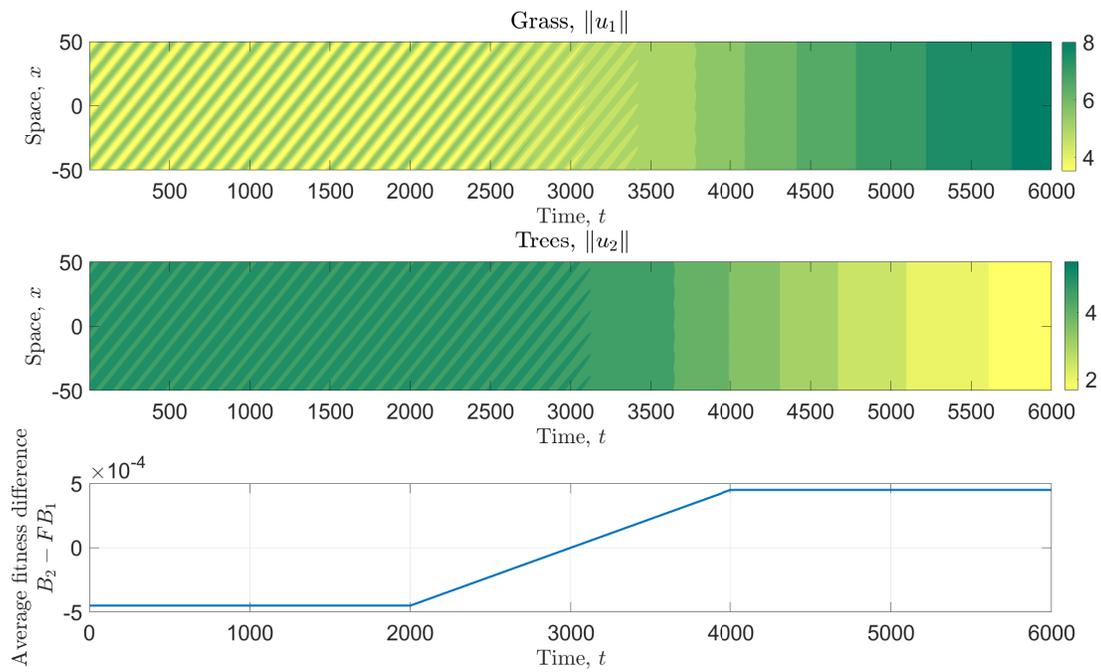}
	\caption{Behaviour of a solution as the average fitness difference changes its sign. This illustration shows the decrease in solution amplitudes of a patterned solution of \eqref{eq: Multispecies slope: Model: nondimensional model} in which both species coexist, as the average fitness difference $B_2-FB_1$ gradually tends to zero from below. At $B_2-FB_1=0$ the solution loses its stability, but no rapid regime shift to a stable single-species state occurs. Instead, both species continue to coexist in a spatially uniform metastable state. The precipitation parameter used in the simulation is $A=4.5$. The average fitness difference is changed by variations in $B_2$ only. } \label{fig: Multispecies Slope: solution under chagnes to avg fitness difference}
\end{figure}

Moreover, the amplitudes of all densities in the coexistence pattern tend to zero as $B_2-FB_1 \rightarrow 0$. In other words, the coexistence pattern approaches a spatially uniform state as the average fitness difference tends to zero. If a coexistence pattern is a stable solution of \eqref{eq: Multispecies slope: Model: nondimensional model} for $B_2-FB_1$ (but see Sec. \ref{sec: Multispecies slope: stability} for more details on stability), then it automatically loses its stability at $B_2-FB_1=0$ as no coexistence equilibrium state is admitted for $B_2-FB_1>0$. The further evolution of such a solution as $B_2-FB_1>0$ was addressed in a previous paper \cite{Eigentler2019Multispecies} for a slightly different model. Those differences (flat ground instead of sloped terrain and an additional term accounting for an asymmetric interspecific competition), however, do not qualitatively affect the relevant results presented here. If the average fitness difference $B_2-FB_1>0$ remains sufficiently small, then coexistence of both plant species occurs as a metastable state. A metastable solution is a long transient state which eventually converges to a stable single-species state. Hence, a coexistence solution of \eqref{eq: Multispecies slope: Model: nondimensional model} remains in a coexistence state for a significant amount of time after it ceases to exist at $B_2-FB_1=0$, provided that $B_2-FB_1 \ll 1$ (see Fig. \ref{fig: Multispecies Slope: solution under chagnes to avg fitness difference}). 

\section{Stability of coexistence pattern}\label{sec: Multispecies slope: stability}

The analysis presented in the previous section provides an insight into the existence of patterned coexistence solutions of \eqref{eq: Multispecies slope: Model: nondimensional model}. Ecologically, however, it is key to gain an understanding of the stability of such solutions. In Sec. \ref{sec: Multispecies slope: origin of coex pattern}, we investigated pattern onset and existence for fixed migration speed $c$. In this section, however, we present stability (and existence) results in the whole $(A,c)$ plane to gain a comprehensive understanding of a pattern's behaviour under changes of the precipitation parameter $A$. 

\begin{figure}
	\centering
	\includegraphics[width=0.7\textwidth]{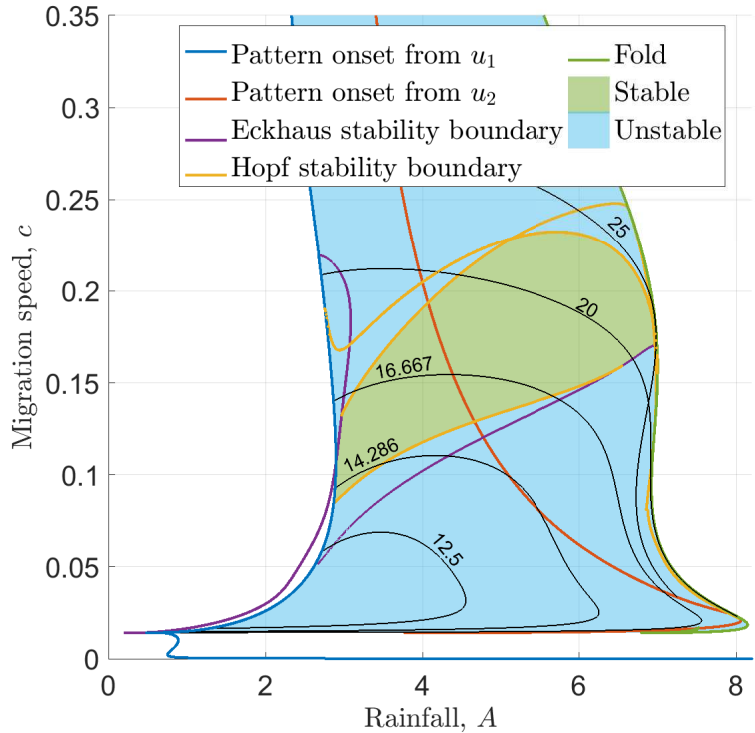}
	\caption{Existence and stability of coexistence patterns. The Busse balloon (parameter region of stable patterns) of patterned solutions of \eqref{eq: Multispecies slope: Model: nondimensional model} in which both species coexist is shown embedded in the existence regions of such solutions in the $(A,c)$ parameter plane. Existence and stability boundaries are computed using the numerical continuation methods outlined in Sec. \ref{sec: Multispecies Slope: spectrum calc}. Wavelength contours are visualised using black solid lines. Note that stability boundaries may extend into regions that are neither marked as stable nor unstable, since biologically irrelevant coexistence patterns with negative densities occur outside the shaded parameter region.}\label{fig: Multispecies Slope: coex stablity ac plane}
\end{figure}

\begin{figure}
	\centering
	\includegraphics[width=0.7\textwidth]{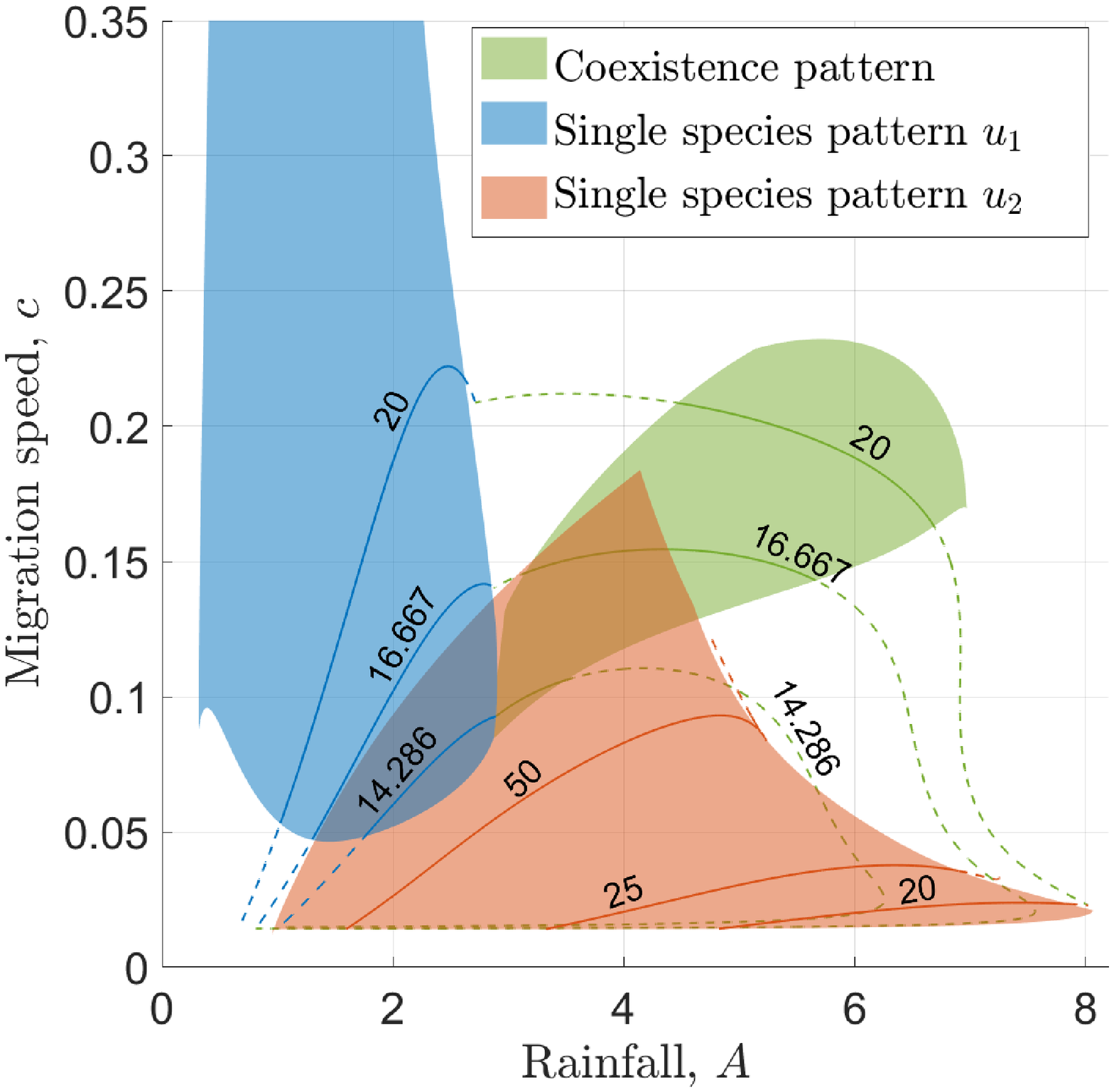}
	\caption{Busse balloons of patterns in the system. This figure visualises the Busse balloons (regions of stable patterns) for the coexistence patterns and both single species patterns that occur as solutions of \eqref{eq: Multispecies slope: Model: nondimensional model}. Wavelength contours are given as solid lines, and their colour indicates the solution type they represent. Solid lines correspond to stable solutions (inside the respective Busse balloon), dashed lines to unstable patterns.}\label{fig: Multispecies Slope: all stablity ac plane}
\end{figure}

The (in)stability of a pattern with given precipitation level $A$ and migration speed $c$ can be determined through a calculation of its essential spectrum. To avoid the computationally expensive calculation of a large number of essential spectra on a fine grid in the $(A,c)$ parameter plane, an extension of the numerical continuation method by Rademacher et al. \cite{Rademacher2007, Sherratt2013a} can be used to trace stability boundaries in parameter space (see Sec. \ref{sec: Multispecies Slope: spectrum calc} and \cite{Rademacher2007, Sherratt2013a} for more details). Stability changes of periodic travelling waves under variations of either the PDE parameters or the migration speed $c$ can be classified into two types \cite{Rademacher2007a}. A stability change of Eckhaus (sideband) type is characterised by a sign change of the curvature of the spectrum at the origin, which is always part of the spectrum due to translation invariance of periodic travelling waves. If instead a pair of folds in the essential spectrum crosses the imaginary axis with nonzero real imaginary part, then the stability change is said to be of Hopf type. Tracing both Eckhaus and Hopf stability boundaries allows us to create a map of stability in the $(A,c)$ plane, often referred to as the \textit{Busse balloon} \cite{Busse1978}. Such a Busse balloon for the coexistence patterns in \eqref{eq: Multispecies slope: Model: nondimensional model} is shown in Fig. \ref{fig: Multispecies Slope: coex stablity ac plane}, where it is embedded into the solution type's existence region. The boundaries for pattern existence in the $(A,c)$ are also obtained by numerical continuations of pattern onset loci and folds along the solution branches. Note that due to the existence of folds in the solution branches of coexistence patterns, an $(A,c)$ pair does not necessarily uniquely define a member of the coexistence pattern solution family. However, our stability analysis indicates that if more than one periodic travelling wave solution of \eqref{eq: Multispecies slope: Model: nondimensional model} exists for a given $(A,c)$ pair, then only a maximum of one of the solutions is stable. For simplicity, we make no distinction between $(A,c)$ pairs that uniquely define a stable pattern and parameter values for which additional unstable patterns exist in our definition of the Busse balloon. Hence, a pair $(A,c)$ is a member of the stability region in the visualisations (Fig. \ref{fig: Multispecies Slope: coex stablity ac plane} and \ref{fig: Multispecies Slope: all stablity ac plane}), even if additional unstable patterns exist.

\begin{figure}
	\begin{overpic}[width=\textwidth]{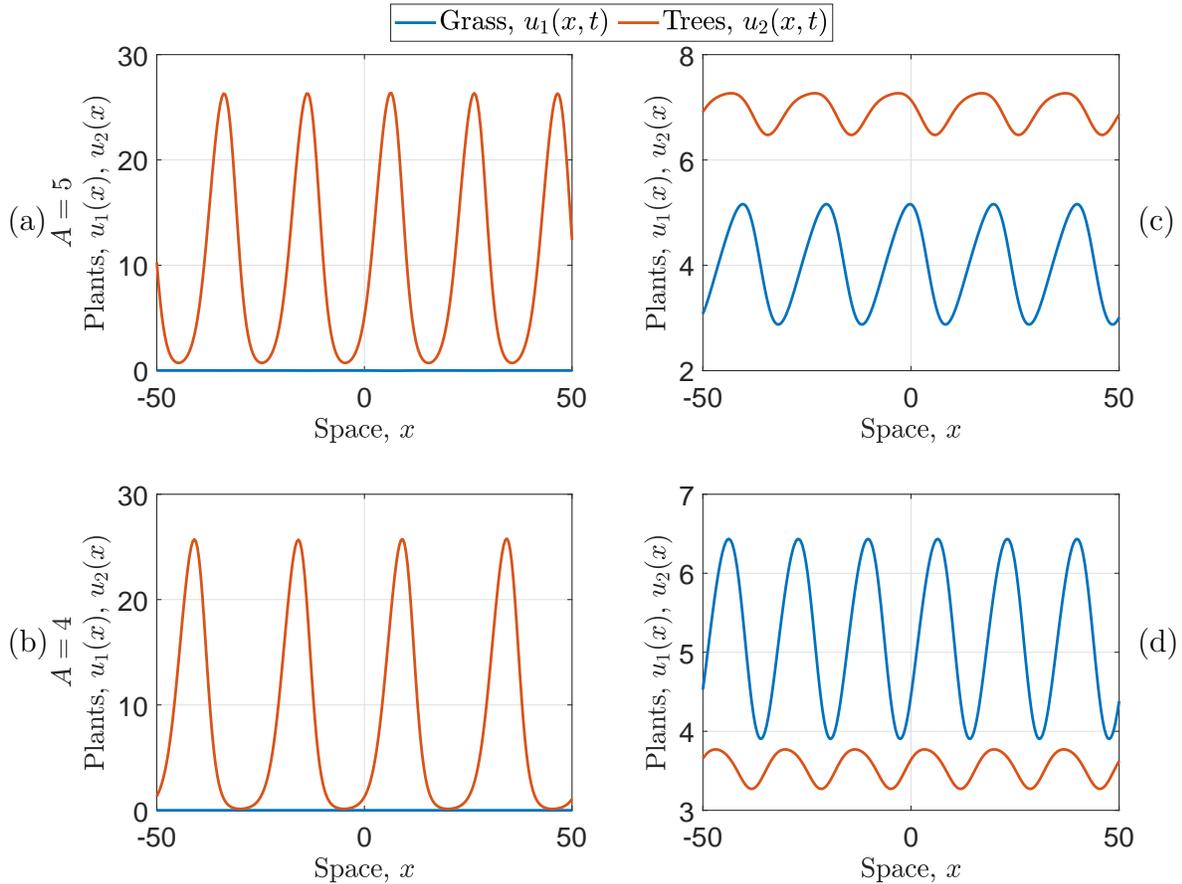}
		\put(1,55){\large (a)}
		\put(1,21){\large (b)}
		\put(92,55){\large (c)}
		\put(92,21){\large (d)}
	\end{overpic}
	\caption{Wavelength changes due to decreasing precipitation. Single-species patterns ((a) and (c)) and multispecies patterns ((b) and (d)) are shown for different precipitation levels to visualise the difference in the wavelength selection at destabilisations due to decreasing rainfall. The first row shows stable patterns for $A=5$. As $A$ is gradually decreased to $A=4$, both patterns lose their stability. The single-species pattern ((a) and (c)) selects a solution of higher wavelength, while the multispecies pattern ((b) and (d)) assumes a pattern of lower wavelength.}\label{fig: Multispecies Slope: wavelength compare sol plots}
\end{figure}

A crucial ecological aspect of patterned solutions of \eqref{eq: Multispecies slope: Model: nondimensional model} is their behaviour as they become unstable due to changes in precipitation. To gain some information on the evolution of a solution under changing rainfall, it is instructive to superimpose wavelength contours on the stability diagram (Fig. \ref{fig: Multispecies Slope: coex stablity ac plane}). Given a stable pattern with given wavelength $L$, the solution follows the wavelength contour if the precipitation parameter is varied, until it reaches a stability boundary. Unlike in previous work on pattern stability in ecological systems \cite{Bennett2019, Dagbovie2014}, we do not observe any qualitative differences between the effects of an instability caused by crossing an Eckhaus boundary and a destabilisation that occurs after a stability boundary of Hopf type is crossed. As the stability boundary is crossed, a new wavelength is selected. Significantly, wavelength selection for the coexistence patterns differs from that of both single species patterns. In the case of a single-species solution, a decrease of precipitation across a stability boundary causes a switch to a higher wavelength pattern, increasing the size of the gaps of bare ground between the vegetation stripes (Fig. \ref{fig: Multispecies Slope: wavelength compare sol plots}(a) and (c)). Conversely, a destabilisation of a coexistence pattern due to decreasing precipitation causes the selection of a shorter wavelength pattern (Fig. \ref{fig: Multispecies Slope: wavelength compare sol plots}(b) and (d)). To understand this difference, it is worth recalling a key difference between the two solution types. The troughs of single species patterns in \eqref{eq: Multispecies slope: Model: nondimensional model} attain values close to $u_i = 0$ and represent alternating areas of high biomass and bare ground regions, while the coexistence patterned solutions oscillate between two nonzero biomass levels, corresponding to a savanna-like state. The selection of a smaller wavelength in the coexistence pattern for decreasing precipitation is associated with a simultaneous decrease of the relative pattern amplitude $(\max u_i - \min u_i)/\|u_i\|$, $i=1,2$ in both species. A reduction in the relative amplitude allows for a compensation of the higher density of vegetation peaks associated with a shorter wavelength to achieve the overall reduction in biomass caused by a decrease in the rainfall parameter $A$.

A second key difference between coexistence and single-species patterns in the system is the patterns' migration speed close to stability boundaries for decreasing precipitation $A$. Single-species patterns experience a decrease in their migration speed $c$ before a destabilisation due to decreasing rainfall occurs. This behaviour is an example of a warning sign of an imminent deterioration of the ecosystem that may be used in predicting regime shifts towards desert in water limited ecosystems \cite{Dakos2011, Gowda2016, Kefi2007, Corrado2014, Rietkerk2004, Saco2018}. Such a reduction in uphill movement is not in general observed for patterned solutions in which both species coexist. Depending on a pattern's wavelength, its migration speed may be increasing or decreasing as the wavelength contour passes through a stability boundary and no clear parametric trends of the uphill movement of the pattern close to a wavelength change can be deduced.

A further significant result obtained from a comparison of stability regions for the three patterned solution types in \eqref{eq: Multispecies slope: Model: nondimensional model} is that key features of the coexistence pattern, such as its wavelength and migration speed, are dominated by and very similar to those of the single-species pattern of the species with faster growth rate (Fig. \ref{fig: Multispecies Slope: all stablity ac plane}). Moreover, if $F$ is sufficiently small, i.e. the species with higher average fitness is growing sufficiently slowly, the Busse balloon of the coexistence patterns and the single-species patterns of the fast-growing species do not overlap, as coexistence patterns are stable for precipitation levels that are higher than those in which the single-species patterns are stable. By contrast, the rainfall levels in which both the coexistence patterns and the single-species patterns of the slow growing species are stable overlap. An important implication of this is a facilitative effect of the fast growing species on the species with a slower growth rate. More precisely, there exist precipitation levels in which, in the absence of a second species, the slow growing species assumes a patterned state with $u_2$ close to zero in the troughs of the pattern, but in which also coexistence patterns are stable. Hence, while $\min u_2 \ll \|u_2\|$ in the absence of a competitor, $\min u_2 \approx \|u_2\|$ if a faster growing species is present in the system. Thus, $u_2$ can attain relatively high densities throughout the whole domain, if it coexists with a faster growing species, instead of appearing as an oscillation between a high density and a biomass level close to zero. This facilitative effect is a case of \textit{ecosystem engineering}, a term coined to describe changes to environmental conditions caused by a species that creates a habitat for other species \cite{Jones1994}.

\section{Phase difference}\label{sec: Multispecies Slope: phase diff}
A striking feature of periodic travelling wave solutions of \eqref{eq: Multispecies slope: Model: nondimensional model} in which both species coexist (see e.g. Fig. \ref{fig: Multispecies slope: numerical simulation}) is a slight phase difference between the oscillations of the two plant species. All model parameters affect the slight shift in the solution profile, but the ratio of the plant species' diffusion coefficients $D$ is found to play the most significant role, as it determines which plant species has higher biomass in the uphill direction.

In the one-species Klausmeier model, the plant density and water density of a patterned solution are typically antiphase (i.e. the peaks in the plant density are at the same locations as the troughs of the water density and vice versa) \cite{Sherratt2011, Kinast2014}. Similarly, in the multispecies model \eqref{eq: Multispecies slope: Model: nondimensional model}, the total plant density $u_1+u_2$ and the water density $w$ are also antiphase. The two components of the total plant density (i.e. the grass density $u_1$ and the tree density $u_2$), however, are slightly out of phase. In the solution shown in Fig. \ref{fig: Multispecies slope: numerical simulation}, for example, local maxima of the grass density $u_1$ are located a short distance in the uphill direction (increasing $x$) away from the corresponding local maxima in the tree density $u_2$.

\begin{figure}
	\begin{minipage}{0.48\textwidth}
		\includegraphics[width=\textwidth]{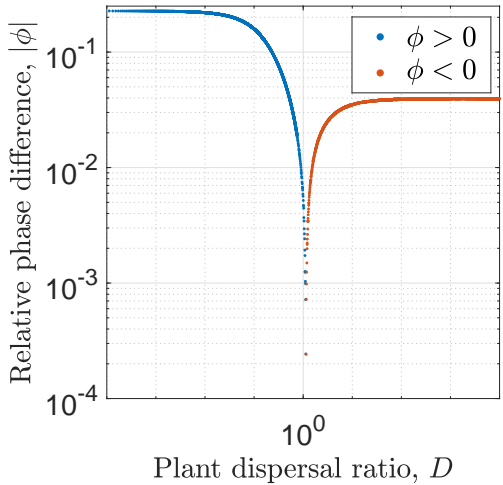}
	\end{minipage}
	\hfill
	\begin{minipage}{0.48\textwidth}
		\caption{Phase difference between the plant species. This figure visualises the absolute value of the relative phase difference in coexistence solutions of \eqref{eq: Multispecies slope: Model: nondimensional model} under changes to the diffusion coefficient $D$, obtained through numerical continuation. The colours indicate the sign of $\phi$, which changes at $D\approx 1$, i.e. when the species' dispersal behaviour is similar. Note the logarithmic scale. The precipitation parameter is $A=4.5$ and the migration speed is set to $c=0.15$.} \label{fig: Multispecies slope: phase difference num cont}
	\end{minipage}	
\end{figure}

Numerical continuation can be used to obtain an insight into the effects of variations in the PDE parameters on the phase difference (Fig. \ref{fig: Multispecies slope: phase difference num cont}). Changes in parameters can have large effects on the period of the patterned solution. We therefore consider the relative phase difference $\phi:= (\arg\max(u_1) - \arg\max(u_2))/L$, where the maxima are taken over one period $0<x<L$, instead of the absolute distance between the two maxima. The tracking of the relative phase difference in solutions obtained through numerical continuation shows that the diffusion coefficient $D$, which describes the ratio of the two plant species' diffusion coefficients, has the most significant effect on the phase difference between the species. If the phase difference $\phi$ is defined as above, then it decreases monotonically with increasing $D$. In particular, it changes its sign close to $D=1$. In other words, if both plant species have similar diffusion coefficients, then their phase difference is small. Note that $\phi=0$ does not necessarily occur at $D=1$, as other model parameters affect the phase difference. The sign change of $\phi$ corresponds to a change in the species which leads the uphill movement of the pattern. Neglecting the phase difference's behaviour in the immediate vicinity of $D=1$, it can be summarised that over one period, the faster dispersing species' maximum and minimum is located a small distance ahead in the uphill direction of the spatial domain.

\section{Discussion}\label{sec: Multispecies Slope: Discussion}

Previous modelling of the savanna biome using nonspatial ODE and impulsive differential equations models (see \cite{Yatat2018} for a review) has successfully identified a range of different mechanisms that stabilise species coexistence based on key differences between grasses and trees. Examples include disturbances that affect species asymmetrically, such as different functional responses in the description of grazing and browsing \cite{Synodinos2015} or variations in the species' susceptibility to fires \cite{Yu2014}; an age structure of trees with different competitive abilities of tree seedlings and adult trees \cite{Baudena2010, DOnofrio2015}; or resource niche separation \cite{Langevelde2003}. Model results presented in this paper suggest that the consideration of spatial interactions in savanna ecosystems can provide an alternative mechanism for species coexistence, as spatial self-organisation principles can facilitate the stable coexistence of grasses and trees in savannas. The novelty of the tree-grass coexistence in model solutions presented in this paper is that both species considered in our multispecies model \eqref{eq: Multispecies slope: Model: nondimensional model} differ only in basic parameters, such as growth rate and mortality rate, and, in particular, satisfy the same single-species model \eqref{eq: Multispecies slope: one species Klausmeier model} for their respective parameter sets.

Solutions of \eqref{eq: Multispecies slope: Model: nondimensional model} in which both species coexist occur, provided that the species with inferior average fitness has a sufficiently large growth rate (Sec. \ref{sec: Multispecies slope: origin of coex pattern}). The average fitness difference $B_2-FB_1$ between the species only depends on the species' growth and mortality rates and determines the system's behaviour in a spatially uniform setting. In particular, $B_2-FB_1=0$ separates the disjoint stability regions of the system's spatially uniform single-species equilibria. The consideration of spatial interactions enables species coexistence as it allows for the capture of effects caused by a positive feedback between local vegetation growth and water redistribution. Patterns of biomass and water densities in the multispecies model \eqref{eq: Multispecies slope: Model: nondimensional model} and the single-species Klausmeier model \eqref{eq: Multispecies slope: one species Klausmeier model} are antiphase (i.e. high water densities in regions of low biomass densities and vice versa). This is due to the depletion of water in regions of high biomass due to the nonlinear dependence of water uptake on the plant densities. The species with faster growth rate (but inferior average fitness) can utilise the higher resource densities in regions of lower biomass through a fast increase in its density in such regions. In the long term, however, it is outcompeted by the species of higher average fitness. This {\color{changes}balance} between local facilitation by the species of higher average fitness and the fast colonisation ability of the species with larger growth rate creates a balance in which coexistence of both species is possible. 

{\color{changes}This result is at odds with those by Durrett and Levin \cite{Durrett1998}, who show that the interplay of local competitiveness and dispersal behaviour it is not sufficient to explain species coexistence in a general competition model, even though it has significant effects on the asymptotic behaviour of the system. A crucial difference between the model by Durrett and Levin and our multispecies ecohydrological model \eqref{eq: Multispecies slope: Model: nondimensional model} is the lack of spatial self-organisation principles in the former. Indeed, if the pattern-inducing feedback is removed from \eqref{eq: Multispecies slope: Model: nondimensional model}, i.e. the infiltration enhancement terms $(u_2+Hu_2)$ are set to unity, no species coexistence occurs in the model. This further emphasises that stable coexistence of the two species is indeed enabled by the spatial heterogeneity in the environmental conditions (water density), which is itself caused by the positive feedback between local plant growth and water redistribution towards high density biomass patches.   }

The model presented in this paper can capture two distinct spatially nonuniform outcomes. Single-species patterns of either species are stable solutions of the system and resemble bands of vegetation that alternate with stripes of bare soil on sloped terrain. In terms of the biomass density, the plant density oscillates between a high level and a level close to zero. By contrast, the second stable patterned solution type features oscillations of both plant species between two non-zero biomass levels. This resembles a savanna state, as plant cover is continuous and no regions of bare soil exist. For typical parameter values of a grass species $u_1$ and a tree species $u_2$, the precipitation intervals of stable single-species tree patterns and stable savanna solutions overlap (Fig. \ref{fig: Multispecies Slope: all stablity ac plane}). This results in the existence of precipitation volumes in which grasses have a local facilitative effect on trees. Under such rainfall regimes and in the absence of a grass species, trees can only attain a patterned state in which tree density oscillates between a high level of biomass and biomass level close to zero. However, if additionally a grass species is considered in the system, trees can coexist with grasses in the whole space domain without the troughs of the oscillations being close to zero. While the total tree biomass decreases if trees coexist with grass, grasses have local facilitative effects on trees as they cause local increases in the tree density. Facilitation occurs due to improvements in environmental conditions. Grasses increase water infiltration into the soil and thus increase resource availability which is utilised by trees, if they are the superior species in a spatially uniform setting. This type of facilitation due to alterations in environmental conditions is referred to as \textit{ecosystem engineering} \cite{Jones1994}. It is well documented in both empirical (e.g. \cite{Pugnaire2001, Moro1997}) and modelling studies (e.g. \cite{Gilad2007a, Meron2007}) that trees can act as ecosystem engineers and facilitate the growth of grass in their vicinity. Our model results suggest that grasses may act as ecosystem engineers too, a mechanism that was established to be the driving force of species coexistence in a model for dryland vegetation patterns by Baudena and Rietkerk \cite{Baudena2013} and backed up by field studies \cite{Anthelme2009, Maestre2003}.

The plant species' diffusion coefficients ratio $D$ has a significant influence on the coexistence solution dynamics. In particular, it quantitatively affects the size of the parameter region giving species coexistence (Sec. \ref{sec: Multispecies slope: origin of coex pattern}). If both species diffuse at the same rate ($D=1$), then coexistence patterns occur if the species with superior average fitness has a slower growth rate. In this case, the inferiority of one species' competitive abilities is balanced by its advantage in its colonisation abilities. The requirement of this crucial balance for species coexistence has already been noted in the early savanna model by Tilman \cite{Tilman1994}. However, in any nonspatial model, spatial spread cannot be distinguished from local growth in the description of a species' colonisation abilities. In the PDE model in this paper, a comparison of local growth rates is only equivalent to a comparison of the plant species' colonisation abilities if the plant species do not differ in their diffusion coefficients. If, however, the inferior competitor in the spatially uniform setting diffuses at a faster rate, then higher growth rates of the superior species are tolerated. Similarly, coexistence patterns also occur if the species of higher average fitness is also superior in its spatial spread, provided that its local growth rate is sufficiently small. 

In the context of species coexistence in vegetation patterns, Nathan et al. \cite{Nathan2013} found that under the assumption that two species decay at an equal rate, coexistence requires a species that is superior in both its competitive (defined by plant growth only) and dispersal abilities, due to a trade-off between spatial spread and local growth. Our results on pattern existence attempt to bridge a gap between the apparent mismatch between the predictions by Tilman \cite{Tilman1994} and Nathan et al. \cite{Nathan2013}. We emphasise that it is essential to consider spatiotemporal models that consider growth and death of plants separately, to gain an understanding of species coexistence. Our results show that, in this case, the complex system dynamics enable species coexistence in different parameter regimes that cover the predictions by both Tilman \cite{Tilman1994} and Nathan et al. \cite{Nathan2013}. In particular, the spatial self-organisation of plants that induces a nonlinear description of biomass growth, renders it insufficient to consider a plant species' competitive ability by one parameter only. The use of the notion of the average fitness of a plant species, comparing its growth rate to its mortality rate, as a measure of its competitive abilities instead, allows to overcome the proposed trade-off between spatial dispersal and local plant growth and enables coexistence of species if the superior competitor diffuses at a slower rate.

Coexistence of species as a model outcome is not limited to the parameter regions discussed above. If no solution with species coexistence occurs in the model, coexistence can occur as a long transient state (towards a stable single-species state), provided that the average fitness difference between the two species is sufficiently small (Fig. \ref{fig: Multispecies Slope: solution under chagnes to avg fitness difference}). We have discussed the concept of metastability as a coexistence mechanism in a previous paper \cite{Eigentler2019Multispecies}, using a model very similar to the multispecies model considered in this paper. The differences between the two models do, however, not qualitatively affect the metastability property. Metastability is characterised by the small (but positive) growth rates of perturbations to a single-species equilibrium that becomes unstable as a competitor is introduced. The size of the growth rate is controlled by the average fitness difference between both species and thus coexistence can occur as a long transient state if the species' competitive abilities are similar, even if coexistence is unstable.

The metastability property is a feature of the spatially uniform model and thus independent of the slope parameter $\nu$ \cite{Eigentler2019Multispecies}. Hence, metastable coexistence also occurs in the system if the terrain is assumed to be flat. The analysis of the stable coexistence states in Sec. \ref{sec: Multispecies slope: origin of coex pattern}-\ref{sec: Multispecies Slope: phase diff}, however, is only valid on a sloped terrain, as the application of the numerical continuation techniques used in the bifurcation and stability analyses rely on the advection term in the equation for the water dynamics. Numerical integration of the PDE system, however, shows that a gradual decrease of the slope parameter to $\nu=0$ does not qualitatively change the behaviour of a stable coexistence state {\color{changes}(in particular the phase difference between the total plant density and the water density)}. By contrast, PDE simulations starting from a randomly perturbed uniform state with the slope parameter fixed to $\nu=0$ yield coexistence solutions in which the pattern wavelength changes frequently. While there is a clear indication that coexistence of species is a potential model outcome  on flat ground, the investigation of the system dynamics would require an application of different analytical tools, which is beyond the scope of this paper.

A distinctive feature of spatially nonuniform solutions of out model is a slight phase difference between both species (Sec. \ref{sec: Multispecies Slope: phase diff}). Such phase differences have been recorded in empirical studies on species coexistence in vegetation bands of semi-arid ecosystems, with grasses reported to be the dominant species in the uphill regions of a stripe, while trees were observed to attain their maximum densities in the central regions of a stripe \cite{Herbes2001}. Our model is unable to reproduce stable solutions that represent species coexistence in vegetation bands, but nevertheless predicts a phase difference between the two species coexisting in a spatially non-uniform savanna state. In particular, in the context of coexistence of grasses and trees (grasses disperse faster than trees), our analysis suggests that the biomass peaks of the herbaceous species are located in the upward direction of the biomass peaks of the woody species. While we are not aware of any data on species-specific biomass distribution in savanna ecosystems, this finding agrees with the empirical data that is available for banded vegetation patterns. \cite{Herbes2001}. In our model, we describe plant spread by diffusion, which is a local mode of dispersal derived from a random walk, and characterise differences in the plant species dispersal behaviour by different diffusion coefficients only. In reality, however, nonlocal processes affect seed dispersal (e.g. \cite{Bullock2017}). Effects of nonlocal plant dispersal on vegetation in semi-arid environments has previously been studied in single-species models \cite{Eigentler2018nonlocalKlausmeier, Bennett2019, Pueyo2008}. A similar approach could be used in an extension of the multispecies model presented in this paper to gain more information on the biomass distribution of both species across a single vegetation stripe.

In this paper, we investigated the facilitative effects of spatial heterogeneities on species coexistence in arid savannas. However, we restricted the extent of spatial heterogeneities to those in the availability of resources caused by a self-organisation principle in the plant populations. In doing so, we neglected potential heterogeneities in the topography of the spatial domain, which may have a significant influence on the ecosystem dynamics \cite{Gandhi2018}. In particular, topographic nonuniformity may alter the dynamics of water flow and thus increase the heterogeneity in the resource availability. Such a promotion of resource niche creation could be exploited in a future model extension to extend the theory on the facilitative effects of spatial interactions in patterned vegetation and arid savannas.

{\color{changes}
	The work presented in this paper not only suggests a novel mechanism for species coexistence in savannas, but also provides insights into other properties of the ecosystem dynamics, such as the slow uphill movement of biomass peaks or the slight phase shift in the species distribution, as discussed above. To test these hypotheses, a comparison with empirical data would be desirable. However, data acquisition for dryland ecosystems is notoriously difficult. Some relevant types of data on dryland ecosystems are available. In particular, Deblauwe et al. \cite{Deblauwe2012} were able to estimate the uphill movement of vegetation stripes by comparing recent satellite images with those taken by spy satellites in the 1960's, but this relied on the clear contrast between vegetation and bare ground - changes in vegetation type within savannas are much more difficult to detect. Data on precipitation (both current and historical \cite{Sherratt2015}) and on elevation (and hence slope) \cite{Sugarbaker2014} are also available. But these are insufficient to provide an effective empirical test of model \eqref{eq: Multispecies slope: Model: nondimensional model}. However, advances in technologies (e.g. image processing) may in the future be utilised to extract more data from satellite images to estimate ecosystem properties of savannas, such as species composition or biomass densities, over large spatial scales. 
}

The study of facilitation between species and mechanisms that promote coexistence is widespread across ecology. In particular, spatial self-organisation has been established as a key element promoting species coexistence in a variety of ecosystems. For example, self-organisation of a macrophyte species in streams enhances environmental conditions through deflection of water and thus facilitates other species through a reduction in environmental stress \cite{Cornacchia2018}. Similarly, self-organised shellfish reefs (in particular mussel beds) are shown to cause a significant increase in species richness and diversity \cite{Christianen2017}. A detailed understanding of facilitative mechanisms caused by spatial self-organisation principles is therefore relevant not only in the vegetation dynamics of semi-arid environments, but also in a wide range of other ecosystems, as it can provide valuable information for restoration and conservation efforts \cite{Cornacchia2018}.

\section{Methods of calculating pattern existence and stability} \label{sec: Multispecies Slope: spectrum calc}
In this section we outline the numerical continuation method by Rademacher et al. \cite{Rademacher2007} to calculate the essential spectrum of a periodic travelling wave and trace stability boundaries of periodic travelling waves in a parameter plane, which we utilised in our bifurcation and stability analysis in Sec. \ref{sec: Multispecies slope: origin of coex pattern} and \ref{sec: Multispecies slope: stability}. We provide an overview of the implementation of the method to \eqref{eq: Multispecies slope: Model: nondimensional model}, but refer the reader to \cite{Rademacher2007, Sherratt2012, Sherratt2013a} for full details. The method described below is implemented using the numerical continuation software AUTO-07p \cite{AUTO}.

\subsection{Single-species pattern existence}

Single-species patterns of both the multispecies model \eqref{eq: Multispecies slope: Model: nondimensional model} and the single-species Klausmeier model \eqref{eq: Multispecies slope: one species Klausmeier model} originate at a Hopf bifurcation and terminate in a homoclinic orbit. Numerical continuation of the Hopf locus in the $(A,c)$ parameter plane is straightforward. The homoclinic orbits, yielding the lower bounds on the precipitation parameter $A$ for pattern existence, may also be calculated by means of numerical continuation. In this context, however, it suffices to approximate homoclinic orbits by periodic travelling waves of large period $L$. Up to some constants in the equilibria and the parameter bounds, identical considerations hold for the second plant species $u_2$, due to the symmetry in the model.

\subsection{Calculation of the essential spectrum} 

The starting point for the calculation of the essential spectrum of a patterned solution of \eqref{eq: Multispecies slope: Model: nondimensional model} is the travelling wave system \eqref{eq: Multispecies slope: Model: tw model}, i.e.
\begin{subequations}\label{eq: Multispecies slope: Model: tw model 1}
	\begin{align}
	f\left(U_1,U_2,W\right)+ c\frac{\dif U_1}{\dif z} +\frac{\dif^2 U_1}{\dif z^2} &=0,\label{eq: Multispecies: Model: tw model u1 1} \\
	g\left(U_1,U_2,W\right) + \frac{\dif U_2}{\dif z} +D\frac{\dif^2 U_2}{\dif z^2}&=0, \label{eq: Multispecies: Model: tw model u2 1}\\
	h\left(U_1,U_2,W\right) + (c+\nu) \frac{\dif W}{\dif z} +d \frac{\dif^2 W}{\dif z^2} &=0,
	\end{align}
\end{subequations}
where
\begin{align*}
f\left(U_1,U_2,W\right) &=  WU_1\left(U_1 + HU_2\right) - B_1 U_1,\\
g\left(U_1,U_2,W\right) &=  FWU_2\left(U_1 + HU_2\right) - B_2 U_2,\\
h\left(U_1,U_2,W\right) &=  A-W - W\left(U_1+U_2\right)\left(U_1 + HU_2\right).
\end{align*}

To determine the essential spectrum, it is further convenient to rewrite the PDE system \eqref{eq: Multispecies slope: Model: nondimensional model} in terms of $z$ and $t$. Denoting $\widehat{u_1}(z,t) = u_1(x,t)$, $\widehat{u_2}(z,t) = u_2(x,t)$ and $\widehat{w}(z,t) = w(x,t)$ thus yields
\begin{subequations} \label{eq: Multispecies slope: PDE in z and t}
	\begin{align}
	\frac{\partial \widehat{u_1}}{\partial t}&= f\left(\widehat{u_1},\widehat{u_2},\widehat{w}\right) + c\frac{\partial \widehat{u_1}}{\partial z} +\frac{\partial^2 \widehat{u_1}}{\partial z^2} ,\\
	\frac{\partial \widehat{u_2}}{\partial t}&= g\left(\widehat{u_1},\widehat{u_2},\widehat{w}\right) + \frac{\partial \widehat{u_2}}{\dif z} +D\frac{\partial^2 \widehat{u_2}}{\partial z^2}, \\
	\frac{\partial \widehat{w}}{\partial t}&= h\left(\widehat{u_1},\widehat{u_2},\widehat{w}\right) + (c+\nu) \frac{\partial \widehat{w}}{\partial z} +d \frac{\partial^2 \widehat{w}}{\partial z^2} .
	\end{align}
\end{subequations}
Given a periodic travelling wave solution $\overline{V}(z) = (\overline{U_1}(z), \overline{U_2}(z), \overline{W}(z))$ of \eqref{eq: Multispecies slope: PDE in z and t} (i.e. a triplet $(\overline{U_1}(z), \overline{U_2}(z), \overline{W}(z))$ that satisfies \eqref{eq: Multispecies slope: Model: tw model 1}), its stability is determined by the behaviour of small perturbations to the periodic travelling wave. Under the assumptions that temporal perturbations to $\overline{V}(z)$ are proportional to $\exp(\lambda t)$, $\lambda\in\C$, i.e. setting $\widehat{v}(z,t) = \overline{V}(z) + \exp(\lambda t) \widetilde{V}(z)$, and linearising \eqref{eq: Multispecies slope: PDE in z and t} about the travelling wave solution $\overline{V}(z)$ yields that the leading order behaviour of perturbations is determined by the eigenvalue problem
\begin{align}
\label{eq: Multispecies slope: ev problem}
\lambda \widetilde{V}(z) = J\widetilde{V}(z) + c\widetilde{V}'(z),
\end{align}
where the prime denotes differentiation with respect to $z$ and $J$ is the Jacobian of the right hand side of \eqref{eq: Multispecies slope: PDE in z and t} with respect to $\widehat{v}$ and its derivatives, i.e.
\begin{align*} 
J = \matthree{\dfrac{\partial f}{\partial \widehat{u_1}} + c\dfrac{\dif}{\dif z} + \dfrac{\dif^2}{\dif z^2}}{\dfrac{\partial f}{\partial \widehat{u_2}}}{\dfrac{\partial f}{\partial \widehat{w}}}
{\dfrac{\partial g}{\partial \widehat{u_1}}}{\dfrac{\partial g}{\partial \widehat{u_2}}+ c\dfrac{\dif}{\dif z} + D\dfrac{\dif^2}{\dif z^2}}{\dfrac{\partial g}{\partial \widehat{w}}}
{\dfrac{\partial h}{\partial \widehat{u_1}}}{\dfrac{\partial h}{\partial \widehat{u_2}}}{\dfrac{\partial h}{\partial \widehat{w}}+ (c+\nu)\dfrac{\dif}{\dif z} + d\dfrac{\dif^2}{\dif z^2}},
\end{align*}
evaluated at the periodic travelling wave solution $\overline{V}$.

The eigenvalue problem \eqref{eq: Multispecies slope: ev problem} is formulated over one period $L$ of the travelling wave solution $\overline{V}(z)$ and needs to be equipped with boundary conditions. By definition, $\overline{V}(0) = \overline{V}(L)$. The eigenfunction $\widetilde{V}(z)$, however, is not necessarily periodic. The amplitude of $\widetilde{V}(z)$ needs to be conserved to prevent growth to $\pm \infty$, but phase shifts are admissible. An appropriate boundary condition thus is
\begin{align}
\label{eq: Multispecies slope: ev problem bc}
\widetilde{V}(0) = \widetilde{V}(L) e^{\gamma i},
\end{align}
for $\gamma \in \R$ which can be derived using Floquet theory \cite{Rademacher2007, Deconinck2006, Sandstede2002}.

The spectral stability of periodic travelling wave solutions $\overline{V}$ can then be determined by finding the set of eigenvalues $\lambda$ for which the eigenvalue problem \eqref{eq: Multispecies slope: ev problem} with boundary condition \eqref{eq: Multispecies slope: ev problem bc} has a nontrivial solution. To do this, it suffices to find the essential spectrum of the periodic travelling wave, as the point spectrum is always empty \cite{Sandstede2002}.

The calculation of the essential spectrum is performed in two stages. First, the special (and simpler) case of periodic boundary conditions (i.e. $\gamma=0$) is considered. This simplification allows for a transformation of the eigenvalue problem \eqref{eq: Multispecies slope: ev problem} into a matrix eigenvalue problem by discretising the domain and approximating the derivatives through finite differences. The matrix eigenvalue problem can be solved by standard means and provides a starting point for a numerical continuation in $\gamma$ to complete the computation of the essential spectrum.

To implement the numerical continuation, it is convenient to rewrite the eigenvalue problem \eqref{eq: Multispecies slope: ev problem} as the first order system
\begin{align*}
\widetilde{\widetilde{V}}(z)' = \left( Y(z) + \lambda X\right) \widetilde{\widetilde{V}}(z), \quad \widetilde{\widetilde{V}}(0) = \widetilde{\widetilde{V}}(L)e^{i\gamma},
\end{align*}
where 
\begin{align*}
Y(z) = \left(\begin{array}{cccccc}
0 & 1 & 0 & 0 & 0 & 0 \\
-\dfrac{\partial f}{\partial U_1} & -c & -\dfrac{\partial f}{\partial U_2} & 0 & -\dfrac{\partial f}{\partial W} & 0 \\ 
0 & 0 & 0 & 1 & 0 & 0 \\
-\dfrac{1}{D} \dfrac{\partial g}{\partial U_1} & 0 & -\dfrac{1}{D} \dfrac{\partial g}{\partial U_2} & -\dfrac{c}{D} & -\dfrac{1}{D} \dfrac{\partial g}{\partial W} & 0 \\
0 & 0 & 0 & 0 & 0 & 1 \\
-\dfrac{1}{d} \dfrac{\partial h}{\partial U_1} & 0 & -\dfrac{1}{d} \dfrac{\partial h}{\partial U_2} & 0 & -\dfrac{1}{d} \dfrac{\partial h}{\partial W} & -\dfrac{c+\nu}{d} 
\end{array}\right),
\end{align*}
evaluated at the periodic travelling wave solution $\overline{V}$ and
\begin{align*}
X = \left(\begin{array}{cccccc}
0 & 0 & 0 & 0 & 0 & 0 \\
1 & 0 & 0 & 0 & 0 & 0 \\ 
0 & 0 & 0 & 0 & 0 & 0 \\
0 & 0 & \dfrac{1}{D} & 0 & 0 & 0 \\
0 & 0 & 0 & 0 & 0 & 0 \\
0 & 0 & 0 & 0 & \dfrac{1}{d} & 0  
\end{array}\right).
\end{align*}
The boundary condition is transformed into a periodic boundary condition by setting $\widetilde{\widetilde{V}}(z) = \exp(i\gamma z/L) \alpha(z)$. Together with the normalisation $z = L\xi$ of the domain, this yields 
\begin{align}
\label{eq: Multispecies slope: ev problem transformed}
\alpha'(\xi) = \left(L\left(Y(\xi) + \lambda X \right) - i \gamma I \right) \alpha(\xi), \quad \alpha(0) = \alpha(1),
\end{align}
where $I$ is the identity matrix. Implementation in AUTO requires separation of real and imaginary parts of \eqref{eq: Multispecies slope: ev problem transformed}. This yields
\begin{subequations}
	\label{eq: Multispecies slope: ev problem real imag}
	\begin{align}
	\Re(\alpha)' &= \left(L \left(Y + \Re(\lambda)X \right)  \right) \Re(\alpha) + \left(\gamma I - L \Im(\lambda)X \right) \Im(\alpha), \label{eq: Multispecies slope: ev problem real imag real} \\
	\Im(\alpha)' &= \left(L \left(Y + \Re(\lambda)X \right)  \right) \Im(\alpha) + \left(-\gamma I + L \Im(\lambda)X \right) \Re(\alpha), \label{eq: Multispecies slope: ev problem real imag imag} \\
	\Re(\alpha(0)) & = \Re(\alpha(1)), \quad  \Im(\alpha(0))  = \Im(\alpha(1)). \label{eq: Multispecies slope: ev problem real imag bc}
	\end{align}
\end{subequations}

The eigenvalue problem \eqref{eq: Multispecies slope: ev problem transformed} is not sufficient to uniquely determine the eigenfunctions $\alpha$. The periodic boundary conditions allow for arbitrary phase shifts. Thus, \eqref{eq: Multispecies slope: ev problem transformed} is supplemented with the phase fixing condition 
\begin{align}
\label{eq: Multispecies slope: phase fix eigenfunction}
\Im \left(\langle \alpha_{\operatorname{old}}, \alpha \rangle \right)  = \int_{0}^{1} \left(\Re\left(\alpha_{\operatorname{old}}\right)\cdot \Im(\alpha) -  \Im\left(\alpha_{\operatorname{old}}\right)\cdot \Re(\alpha) \right) \dif \xi = 0,
\end{align}
where $\alpha_{\operatorname{old}}$ is the eigenfunction $\alpha$ at any previous step of the numerical continuation or the initial eigenfunction from which the continuation is started, and the inner product is defined by
\begin{align*}
\langle \alpha_1, \alpha_2 \rangle = \int_{0}^1 \alpha_1 \cdot \alpha_2^\ast \dif \xi,
\end{align*}
where the asterisk denotes the complex conjugation. Further, the eigenfunction is normalised by imposing the integral condition
\begin{align}
\label{eq: Multispecies slope: normalise eigenfunction}
\langle \alpha, \alpha \rangle = \int_{0}^1 \left( \Re(\alpha) \cdot \Re(\alpha) + \Im(\alpha) \cdot \Im(\alpha) \right) \dif \xi = 1.
\end{align}
Similar to the phase fixing condition \eqref{eq: Multispecies slope: phase fix eigenfunction} for the eigenfunction $\alpha$, also the periodic travelling wave solution $V = (U_1,U_1',U_2,U_2',W,W')$ of \eqref{eq: Multispecies slope: Model: tw model 1} with periodic boundary conditions requires a phase fixing condition to prevent arbitrary translations in $z$. The appropriate integral condition is
\begin{align}
\label{eq: Multispecies slope: phase fix ptw}
\int_{0}^1 V_{\operatorname{old}}' \cdot \left(V_{\operatorname{old}} - V\right) \dif z = 0.
\end{align}

Given a solution of the eigenvalue problem \eqref{eq: Multispecies slope: ev problem} with periodic boundary conditions (i.e. $\gamma = 0$), the full essential spectrum can then be found by continuing the travelling wave equation \eqref{eq: Multispecies slope: Model: tw model 1} with periodic boundary conditions and the eigenfunction equation \eqref{eq: Multispecies slope: ev problem real imag} with the integral constraints \eqref{eq: Multispecies slope: phase fix eigenfunction}, \eqref{eq: Multispecies slope: normalise eigenfunction} and \eqref{eq: Multispecies slope: phase fix ptw}, starting from each of the eigenvalues $\lambda$ and corresponding eigenfunctions $\alpha$ obtained from the matrix eigenvalue problem for $\gamma = 0$. The principal continuation parameter is $0<\gamma<2\pi$, while $\Re(\lambda)$, $\Im(\lambda)$ and $L$ are chosen as secondary continuation parameters. In practise, not the whole essential spectrum needs to be computed to determine the spectral stability of a given periodic travelling wave solution. It is sufficient to perform the numerical continuation starting only from the, say 20, largest eigenvalues obtained form the matrix eigenvalue problem for $\gamma = 0$.

\subsection{Numerical continuation of stability boundaries} 
The method described in the previous section allows for the calculation of the essential spectrum of a periodic travelling wave solution for a set of given parameters. The algorithm can further be extended to trace stability boundaries of periodic travelling waves in a parameter plane, such as $(A,c)$. Full details of this algorithm are found in \cite{Rademacher2007, Sherratt2013a}. 

To locate and trace stability boundaries, derivatives of the eigenfunctions $\alpha$ with respect to $\gamma$ are required. Implicit differentiation of \eqref{eq: Multispecies slope: ev problem transformed} with respect to $\gamma$ gives
\begin{align}
\label{eq: Multispecies slope: ev problem gamma}
\alpha_\gamma' = \left(L \left(Y+\lambda X\right) - i \gamma I \right) \alpha_\gamma + \left(L\lambda_\gamma X - iI \right) \alpha, \quad \alpha_\gamma(0) = \alpha_\gamma(1),
\end{align}
where the prime denotes derivatives with respect to $\xi$ and the subscript $\gamma$ derivatives with respect to $\gamma$. Further implicit differentiation yields
\begin{align}
\label{eq: Multispecies slope: ev problem gamma2}
\alpha_{\gamma\gamma}' = \left(L \left(Y+\lambda X \right) - i\gamma I \right)\alpha_{\gamma\gamma} + 2\left(L\lambda_\gamma X - iI \right) \alpha_\gamma + L \lambda_{\gamma\gamma} X \alpha , \quad \alpha_{\gamma\gamma}(0) = \alpha_{\gamma\gamma}(1).
\end{align}
As previously discussed, implementation in AUTO requires separation of real and imaginary parts. This yields
\begin{subequations}
	\label{eq: Multispecies slope: ev problem gamma real imag}
	\begin{align}\begin{split} \label{eq: Multispecies slope: ev problem gamma real imag real}
	\Re \left(\alpha_\gamma'\right) & = L \left(Y+\Re(\lambda)X\right) \Re\left(\alpha_\gamma\right) + \left( - L\Im(\lambda) X + \gamma I \right) \Im \left(\alpha_\gamma\right) \\ 
	&+ L\Re\left(\lambda_\gamma\right)X\Re(\alpha) + \left(-L\Im\left(\lambda_\gamma\right) X + I  \right) \Im(\alpha), \end{split} \\  \begin{split} \label{eq: Multispecies slope: ev problem gamma real imag imag}
	\Im \left(\alpha_\gamma'\right) & =\left(  L\Im(\lambda) X - \gamma I \right) \Re\left(\alpha_\gamma\right) +   L \left(Y+\Re(\lambda)X\right)\Im \left(\alpha_\gamma\right) \\ 
	&+ \left(L\Im\left(\lambda_\gamma\right) X - I  \right)\Re(\alpha) +  L\Re\left(\lambda_\gamma\right)X \Im(\alpha), \end{split} \\ \label{eq: Multispecies slope: ev problem gamma real imag bc}
	\Re\left(\alpha_\gamma(0)\right) &= \Re\left(\alpha_\gamma(1)\right), \quad \Im\left(\alpha_\gamma(0)\right) = \Im\left(\alpha_\gamma(1)\right),  
	\end{align}	
\end{subequations}
and 
\begin{subequations}
	\label{eq: Multispecies slope: ev problem gamma2 real imag}
	\begin{align}\begin{split} \label{eq: Multispecies slope: ev problem gamma2 real imag real}
	\Re \left(\alpha_{\gamma\gamma}'\right) & = L\left(Y+\Re(\lambda X)\right)  \Re\left(\alpha_{\gamma\gamma}\right) + \left(-L\Im(\lambda)X + \gamma I \right) \Im\left(\alpha_{\gamma\gamma}\right) \\ 
	&+ 2L\Re\left(\lambda_\gamma\right)X \Re\left(\alpha_\gamma\right) + 2\left(-L\Im\left(\lambda_\gamma\right)X + I \right) \Im \left(\alpha_\gamma\right) \\ 
	&+L \Re\left(\lambda_{\gamma\gamma}\right)X \Re(\alpha) -L \Im\left(\lambda_{\gamma\gamma}\right)X \Im(\alpha), \end{split} \\  \begin{split} \label{eq: Multispecies slope: ev problem gamma2 real imag imag}
	\Im \left(\alpha_{\gamma\gamma}'\right) & = \left(L\Im(\lambda)X - \gamma I \right) \Re\left(\alpha_{\gamma\gamma}\right) + L\left(Y+\Re(\lambda X)\right) \Im\left(\alpha_{\gamma\gamma}\right) \\ 
	&+2\left(L\Im\left(\lambda_\gamma\right)X - I \right) \Re\left(\alpha_\gamma\right) +  2L\Re\left(\lambda_\gamma\right)X \Im \left(\alpha_\gamma\right) \\ 
	&+ L \Im\left(\lambda_{\gamma\gamma}\right)X \Re(\alpha) + L \Re\left(\lambda_{\gamma\gamma}\right)X  \Im(\alpha), \end{split} \\ \label{eq: Multispecies slope: ev problem gamma2 real imag bc}
	\Re\left(\alpha_{\gamma\gamma}(0)\right) &= \Re\left(\alpha_{\gamma\gamma}(1)\right), \quad \Im\left(\alpha_{\gamma\gamma}\right) = \Im\left(\alpha_{\gamma\gamma}(1)\right),  
	\end{align}	
\end{subequations}
respectively.

Equations \eqref{eq: Multispecies slope: ev problem gamma} and \eqref{eq: Multispecies slope: ev problem gamma2} cannot determine the derivatives $\alpha_\gamma$ and $\alpha_{\gamma\gamma}$ uniquely, as they may contain components in the nullspace of \eqref{eq: Multispecies slope: ev problem transformed}. Hence, they are equipped with integral conditions given by
\begin{align}
\label{eq: Multispecies slope: int cond alpha gamma}
\langle \alpha, \alpha_\gamma \rangle = 0
\end{align}
and
\begin{align}
\label{eq: Multispecies slope: int cond alpha gamma2}
\langle \alpha, \alpha_{\gamma\gamma} \rangle = 0
\end{align}

A stability change of Eckhaus (sideband) type is detected through a numerical continuation of the travelling wave equation \eqref{eq: Multispecies slope: Model: tw model 1}, the eigenfunction equation \eqref{eq: Multispecies slope: ev problem real imag}, the imaginary part of the eigenvalue equation differentiated with respect to $\gamma$ \eqref{eq: Multispecies slope: ev problem gamma real imag imag} and the real part of the eigenvalue equation differentiated twice with respect to $\gamma$ \eqref{eq: Multispecies slope: ev problem gamma2 real imag real} with the corresponding boundary and integral conditions. The continuation is started at the eigenvalue $\lambda = 0$ and its corresponding eigenfunction obtained from the matrix eigenvalue problem that is solved in the initial stage of the algorithm. The principal continuation parameter is the migration speed $c$ (or the PDE parameter $A$), and the five secondary continuation parameters must include $\Re(\lambda_{\gamma\gamma})$. If a locus with $\Re(\lambda_{\gamma\gamma}) = 0$ is found, a stability change of Eckhaus type is detected. The secondary continuation parameter $\Re(\lambda_{\gamma\gamma})$ is then replaced by the PDE parameter $A$ (or the migration speed $c$) to trace out the stability boundary in the $(A,c)$ parameter plane.

The continuation of a stability change of Hopf type follows the same idea, but contains some caveats. First, a fold in the spectrum is detected by a numerical continuation of the travelling wave equation \eqref{eq: Multispecies slope: Model: tw model 1}, the eigenfunction equation \eqref{eq: Multispecies slope: ev problem real imag} and both the real and imaginary parts of the eigenvalue equation differentiated with respect to $\gamma$ \eqref{eq: Multispecies slope: ev problem gamma real imag} with the corresponding boundary and integral conditions. The spectrum may contain many folds, but only the fold with largest real part is of interest and the continuation must start sufficiently close to that fold. The principal continuation parameter is $\gamma$ and the five secondary continuation parameters must include $\Re(\lambda_\gamma)$. A fold in the spectrum is located, when a zero of $\Re(\lambda_\gamma)$ is found. The zero of $\Re(\lambda_\gamma)$ is subsequently fixed and the migration speed $c$ (or the PDE parameter $A$) is then chosen as the principal continuation parameter. The equations are continued in this parameter until a zero of $\Re(\lambda)$, which needs to be one of the secondary continuation parameters, is found. This corresponds to a stability change of Hopf type. Finally, $\Re(\lambda)$ is replaced as a secondary continuation parameter by the PDE parameter $A$ (or the migration speed $c$) to trace out the locus of the stability change of Hopf type in the $(A,c)$ plane.

\section*{Acknowledgements}
Lukas Eigentler was supported by The Maxwell Institute Graduate School in Analysis and its Applications, a Centre for Doctoral Training funded by the UK Engineering and Physical Sciences Research Council (grant EP/L016508/01), the Scottish Funding Council, Heriot-Watt University and the University of Edinburgh.

\clearpage

\bibliography{bibliography.bib}

\end{document}